\documentclass[a4paper,11pt,usenames,dvipsnames]{article}
\pdfoutput=1 

\usepackage{jheppub}
\usepackage[T1]{fontenc} 
\usepackage{amsmath,amssymb,amsfonts,amsthm}
\usepackage{graphicx}
\usepackage[usenames,dvipsnames]{color}
\usepackage{enumitem}
\usepackage{verbatim}
\usepackage[percent]{overpic}
\usepackage{rotating}
\usepackage{hyperref}
\usepackage{array}
\usepackage{mathtools}
\usepackage{lmodern}
\usepackage[normalem]{ulem}
\usepackage{braket}
\usepackage{tensor}
\usepackage{dsfont}
\usepackage{bm}
\usepackage{tikz}
\newcommand{\dvol}[1]{\text{dvol}_{#1}}
\newcommand{\sol}{\text{Sol}}

\usepackage{mathrsfs}

\usetikzlibrary{decorations.pathmorphing}
\usepackage{CJKutf8}

\usetikzlibrary{positioning}
\usetikzlibrary{calc,through,backgrounds}

\theoremstyle{definition}
\newtheorem{definition}{Definition}[section]
\newtheorem{theorem}{Theorem}[section]

\newcommand{\kako}[1]{\left( #1 \right)}
\newcommand{\kagikako}[1]{\left[ #1 \right]}
\newcommand{\ts}[1]{ _{\text{#1}} }
\newcommand{\Bigkako}[1]{\Big( #1 \Big)}

\newcommand{\Bigkagikako}[1]{\Big[ #1 \Big]}

\newcommand{\erfi}{\text{erfi}}

\allowdisplaybreaks

\DeclareMathOperator{\Tr}{Tr}
\newcommand{\R}{\mathbb{R}}
\newcommand{\C}{\mathbb{C}}
\newcommand{\mfd}{\mathcal{M}}
\newcommand{\dd}{\text{d}}
\newcommand{\bk}{{\bm{k}}}
\newcommand{\bx}{{\bm{x}}}

\newcommand{\id}{\mathds{1}}
\newcommand{\sx}{\mathsf{x}}

\newcommand{\ii}{\mathsf{i}}

\newcommand{\kk}{|\bm{k}|}

\title{Relativistic quantum Otto engine: Instant work extraction from a quantum field}

\author[a]{Kensuke Gallock-Yoshimura}

\affiliation[a]{Department of Physics, Kyushu University, \\
744 Motooka, Nishi-Ku, Fukuoka 819-0395, Japan}

\emailAdd{gallockyoshimura.kensuke@phys.kyushu-u.ac.jp}

\abstract{In this study, we carry out a non-perturbative approach to a quantum Otto engine, employing an Unruh-DeWitt particle detector to extract work from a quantum Klein-Gordon field in an arbitrary globally hyperbolic curved spacetime. 
We broaden the scope by considering the field in any quasi-free state, which includes vacuum, thermal, and squeezed states. 
A key aspect of our method is the instantaneous interaction between the detector and the field, which enables a thorough non-perturbative analysis. 
We demonstrate that the detector can successfully extract positive work from the quantum Otto cycle, even when two isochoric processes occur instantaneously, provided the detector in the second isochoric process receives a signal from the first interaction. 
This signaling allows the detector to release heat into the field, thereby the thermodynamic cycle is completed. 
As a demonstration, we consider a detector at rest in flat spacetime and compute the work extracted from the Minkowski vacuum state. 
}

\begin{document} 
\maketitle
\flushbottom

\section{Introduction}

In the realm of quantum thermodynamics, the exploration of heat engines driven by quantum systems has gained significant attention over the past few decades. 
A particularly important example of quantum heat engines is the \textit{quantum Otto engine} (QOE), where a quantum working substance such as a qubit extracts work from heat baths by operating through two isochoric and adiabatic processes \cite{Scovil.heat.engine.maser, Feldmann.QOE.2000, Kieu.secondlaw.demon.otto, Kieu.heat.engine.2006, Rostovtsev.Otto.2003, Quan.multilevel.engine.2005, Quan.multilevel.heat.engine}. 
The QOE is particularly significant as it enables separate evaluations of heat and work \cite{Bhattacharjee2021}.

While quantum heat engines have been extensively studied within the domain of quantum mechanics, their exploration in the context of relativistic quantum mechanics \cite{Munoz.QHE.relativistic.dirac.2012, Pena.RQM.engine.2016, Purwanto.RQHE.2016, YIN201858, Chattopadhyay.relativistic.uncertainty.2019, Myers.QOE.relativistic.2021, Chattopadhyay.noncommutative.thermo.2021, Sukamto.RQHE.minimal.length.2023} and quantum field theory remains relatively unexplored. 
Recent investigations \cite{UnruhOttoEngine, Finn.UnruhOtto} have delved into QOEs where the working substance is an Unruh-DeWitt (UDW) particle detector \cite{Unruh1979evaporation, DeWitt1979} interacting with quantum scalar and spinor fields. 
In their QOE, the field is assumed to be in the Minkowski vacuum state, and work is extracted from a thermal bath induced by the Unruh effect \cite{Unruh1979evaporation} experienced by a linearly accelerating detector. 
Such a QOE driven by the Unruh effect is particularly called the Unruh QOE. 
See also \cite{Xu.UnruhOtto.degenerate, Kane.entangled.Unruh.Otto, Barman.entangled.UnruhOtto, Mukherjee.UnruhOtto.boundary} for subsequent studies. 
The Unruh QOE concept extends to scenarios where a UDW detector follows an arbitrary timelike trajectory in curved spacetime \cite{Gallock2023Otto}. 
We will refer to quantum heat engines that include relativistic effects as `relativistic quantum heat engines.'

In both traditional and Unruh QOEs, the working substance is typically assumed to experience thermality, reflecting the engines' roots in thermodynamics. 
However, it raises the question of what occurs if the baths are in a non-thermal state \cite{Abah.nonequilibrium.2014}. 
Remarkably, non-thermal baths, such as quantum coherent \cite{Scully.QHE.coherent.2003} and squeezed thermal baths \cite{Huang.squeezing.work, Rossnagel.Nanoscale.squeezed.QHE.2014, Manzano.squeezed.Otto.2016, Niedenzu.squeezed.QOE.2016, Klaers.squeezed.thermal.QOE.2017}, could offer significant advantages. 
Therefore, a generalized version of QOE, where the quantum state of reservoirs is not restricted, is of great interest. 
In addition, the absence of a thermality requirement means that a working substance need not interact for a long time with a bath. 
As a result, the time dependence of this interaction can also be arbitrary.

In this context, Ref.~\cite{Gallock2023Otto} provides a pertinent example. 
The study extends the Unruh QOE to a more general relativistic QOE, where the background geometry, the state of the field (reservoir), the detector's trajectory, and the time dependence of the interaction are not specified. 
Using perturbation theory, the study has shown that the positive work condition is determined by the effective temperature observed by the qubit.

Building upon these ideas, this paper examines the relativistic QOE under the conditions of instantaneous interaction, characterized by delta-coupling. 
This approach contrasts with the traditionally assumed prolonged interactions in QOE literature and allows us to apply a non-perturbative method to our analysis. 
Our model assumes an arbitrary globally hyperbolic spacetime and permits the UDW detector to follow any timelike trajectory. 
Furthermore, the state of a quantum scalar field is chosen to be one of the quasi-free states such as vacuum and thermal states in quantum field theory in curved spacetime (QFTCS). 
Since QFTCS is known to have infinitely many unitarily inequivalent Hilbert space representations, we adopt an algebraic quantum field theoretic approach to treat all representations on an equal footing.

Our findings reveal that a relativistic QOE with delta-coupling can successfully extract work from a quantum scalar field, \emph{provided the UDW detector is capable of signal exchange through the quantum field across time}. 
This means work extraction is contingent on the detector receiving a signal during the second isochoric process, which was initially sent during the first interaction. 
In other words, if such signal exchange is not feasible then it is impossible to extract work irrespective of the state of the field, background geometry, or the trajectory of the detector. 
As an example, we show that a detector at rest in Minkowski spacetime can extract work from the Minkowski vacuum due to this signaling effect.

Throughout this paper, we use natural units $\hbar = k\ts{B}= c=1$ and the mostly plus metric convention, and denote a spacetime point by $\sx\equiv (t,\bx)$.

\section{Quantum Otto engine: Review}\label{sec:QOE review}
\begin{figure}[t]
\centering
\includegraphics[width=\linewidth]{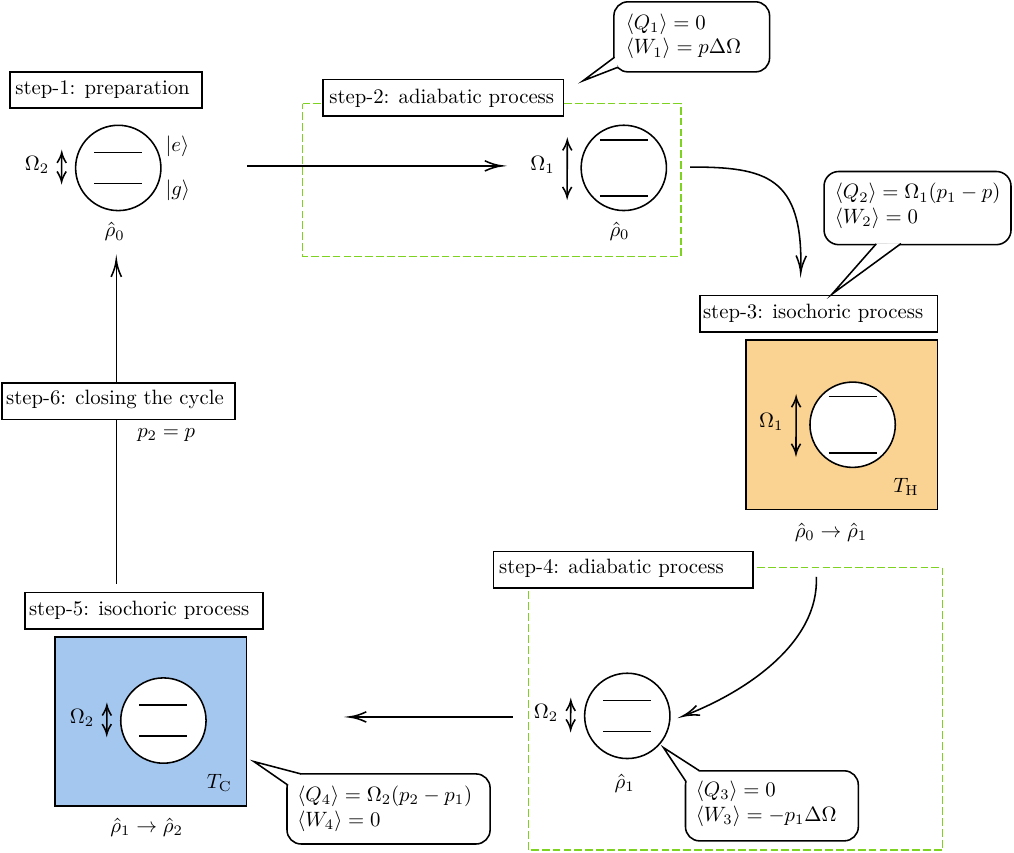}
\caption{A traditional QOE using a qubit interacting with thermal baths at temperatures $T\ts{H}$ and $T\ts{C}$.}
\label{fig:Otto cycle}
\end{figure}

We first review the quantum Otto cycle following the paper by Kieu \cite{Kieu.secondlaw.demon.otto}. 
Here, the working substance is a qubit interacting with heat baths at temperatures $T\ts{H}$ and $T\ts{C}$. 
Figure~\ref{fig:Otto cycle} depicts each step in the cycle. 
It should be noted that the assumption regarding the baths being in a thermal state in this context will be replaced with an arbitrary quasi-free state in Section~\ref{sec: results}.

\begin{enumerate}
 \item Prepare a qubit with energy gap $\Omega_2$ in an initial state $\hat\rho_0= p \ket{e}\bra{e} + (1-p) \ket{g} \bra{g}$ with $p \in [0,1]$. \label{step:step1}
 \item \label{step:step2}An adiabatic process. 
 The qubit adiabatically (i.e., no heat exchange) expands the energy gap from $\Omega_2$ to $\Omega_1$, where $\Omega_1 > \Omega_2$. 
 The Hamiltonian of the qubit during this process is $\hat H(t)=\Omega(t) \ket{e}\bra{e}$ and so the work done on the qubit is 
 \begin{align}
     \braket{W_1}
     &=
        \int \dd t\,
        \Tr \kagikako{
            \hat\rho_0 \dfrac{ \dd \hat H(t) }{\dd t}
        }
    =
        p \Delta \Omega\,,
 \end{align}
 where $\Delta \Omega \coloneqq \Omega_1 - \Omega_2$. 
 The quantum adiabatic theorem tells us that the state of the qubit, $\hat \rho_0$, remains the same during this process. 
 Therefore, the heat flow from the environment to the qubit is zero: 
 \begin{align*}
     \braket{Q_1}
     &=
        \int \dd t\,
        \Tr 
        \kagikako{
            \dfrac{\dd \hat \rho_0}{\dd t} \hat H(t)
        }
    =0\,.
 \end{align*}
 \item \label{step:step3}An isochoric process. 
 The qubit with a fixed energy gap $\Omega_1$ interacts with a bath at temperature $T\ts{H}$. 
 Since the qubit's Hamiltonian is time-independent the work done on the qubit is zero: $\braket{W_2}=0$. 
 Meanwhile the state of the qubit changes from $\hat \rho_0$ to $\hat \rho_1= p_1 \ket{e}\bra{e} + (1- p_1) \ket{g} \bra{g}$. 
 The first law of thermodynamics gives the absorbed heat: 
 \begin{align}
     \braket{Q_2}
     &=
        \Tr[\hat \rho_1 \hat H] - \Tr[\hat \rho_0 \hat H]
    =
        \Omega_1 (p_1-p)\,.
 \end{align}
 \item \label{step:step4}Another adiabatic process. 
 Isolate the qubit from the heat bath and contract the energy gap adiabatically from $\Omega_1$ to $\Omega_2$. 
 The state remains to be $\hat\rho_1$ throughout this process. 
 As in step-\ref{step:step2}, the heat and work are
 \begin{align}
     \braket{Q_3}
     &=
        0\,, \\
    \braket{W_3}
    &=
        -p_1 \Delta \Omega\,.
 \end{align}
 \item \label{step:step5}Another isochoric process. 
 We put the qubit in a colder bath at temperature $T\ts{C} (< T\ts{H})$. 
 At the end of the process, the state becomes $\hat\rho_2=p_2 \ket{e}\bra{e} + (1 - p_2) \ket{g}\bra{g}$, and the heat and work are
 \begin{align}
     \braket{Q_4}
     &=
        \Omega_2 (p_2 - p_1)\,,\\
    \braket{W_4}
    &=
        0\,.
 \end{align}
 \item \label{step:step6}Completing a thermodynamic cycle. 
 The final state of the qubit should coincide with the initial one. 
 Thus we impose $p_2 = p$. 
 We refer to this as a cyclicity condition. 
\end{enumerate}

From the quantum Otto cycle given above, the total extracted work $\braket{W\ts{ext}}$ and absorbed heat $\braket{Q}$ read
 \begin{align}
     \braket{W\ts{ext}}
     &=
        -\sum_{j=1}^4 \braket{W_j}
    =
        (p_1-p) \Delta \Omega
        (=\braket{Q})\,. \label{eq:total extracted work} 
\end{align}
Observe that the first law of thermodynamics is satisfied: $\braket{Q} = \braket{W\ts{ext}}$. 
The positive work condition (assuming $\Omega_1 > \Omega_2$) reads $p_1-p>0$, which turns out to be $T\ts{C}/\Omega_1 < T\ts{H}/\Omega_2$ \cite{Feldmann.QOE.2000, Kieu.secondlaw.demon.otto}. 
Moreover, the efficiency of the QOE is $\eta\ts{O} \coloneqq \braket{W\ts{ext}}/Q_2 =1-\Omega_1/\Omega_2$.

We again note that the baths at temperatures $T\ts{H}$ and $T\ts{C}$ will be replaced with a quantum field in an arbitrary quasi-free state in Section~\ref{sec: results}. 
Moreover, we will employ instantaneous isochoric processes in our analysis. 
Thus, our relativistic QOE is characterized as an engine with: 
(i) a four-stroke cycle comprising two isochoric and two adiabatic processes, where the isochoric processes may have arbitrary time-dependence (including instantaneous interactions) and the adiabatic processes are assumed ideal, ensuring no heat exchange occurs; 
(ii) a complete thermodynamic cycle that returns the qubit to its initial state.

\section{Quantum field theory in curved spacetime}\label{sec:QFTCS}

\begin{figure}[t]
\centering
\includegraphics[width=\linewidth]{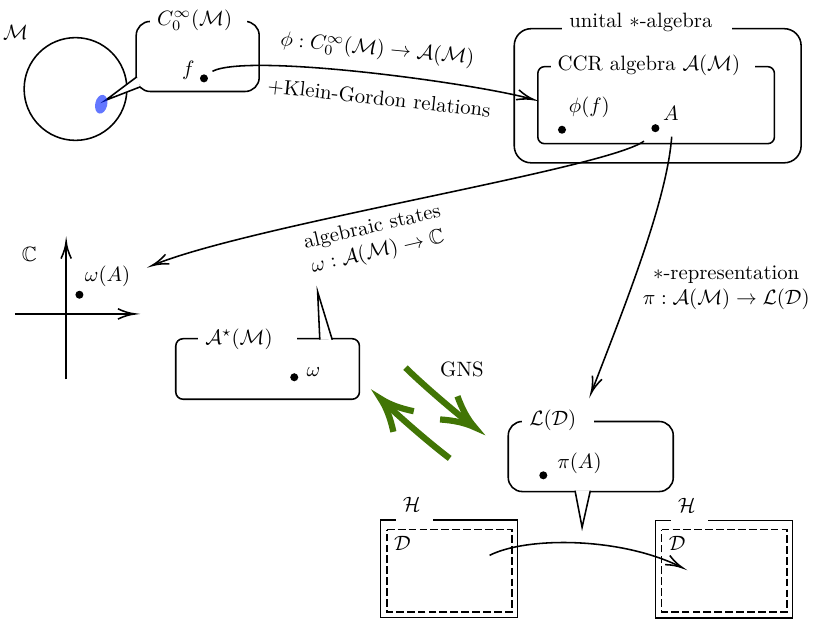}
\caption{A schematic diagram showing the relationship among mathematical concepts. 
A manifold $\mfd$ is depicted as a spherical object, with a compact smooth function $f$ indicated in blue. 
Black arrows represent mathematical mappings from one space to another, e.g., $\phi: C_0^\infty (\mfd) \to \mathcal{A}(\mfd)$. 
$\mathcal{A}^\star(\mfd)$ and $\mathcal{L}(\mathcal{D})$ are linear spaces of algebraic states and linear operators, respectively, where $\mathcal{D} \subset \mathcal{H}$ is a dense subspace in Hilbert space $\mathcal{H}$ shown as a dashed box. 
The correspondence between a representation $(\mathcal{H}, \mathcal{D}, \pi)$ and an algebraic state $\omega$ is indicated by green arrows, with one such relationship being derived from the GNS theorem.
}
\label{fig:AQFT1}
\end{figure}

In this section, we provide a comprehensive review of algebraic quantum field theory (AQFT) based on \cite{Wald:QFTCS, KayWald1991theorems, khavkine2015algebraic, Hollands.and.Wald2015, fewster2020algebraic} as well as \cite{Tjoa.RQC, Tjoa.nonperturbative.gaussian}. 
Employing AQFT is crucial to address the inherent challenges of QFTCS, giving rise to many unitarily inequivalent representations. 
Unlike the conventional Hilbert space approach in QFTCS, in which a specific representation is chosen initially, AQFT treats field observables as elements of an abstract algebra rather than as operators on a Hilbert space, thereby postponing the selection of a particular representation. 
This approach enables us to treat all the unitarily inequivalent Hilbert spaces on the same footing \cite{Wald:QFTCS}. 
Figure~\ref{fig:AQFT1} schematically depicts the relationships among the concepts in AQFT.

\subsection{Algebraic quantum field theory}\label{subsec:AQFT}

Let $(\mfd, g)$ be an $(n+1)$-dimensional globally hyperbolic spacetime. 
In the traditional Hilbert space approach to QFTCS, a free quantum scalar field $\hat \phi$ in this spacetime is considered to be an operator on a Hilbert space satisfying the Klein-Gordon equation: 
\begin{align}
    P\hat \phi=0\,,
    \quad
    P\coloneqq
        \nabla_a \nabla^a - m^2 - \xi R\,,\label{eq:KG eq}
\end{align}
where $\nabla$ is the Levi-Civita connection with respect to the metric $g$, $m\geq 0$ is the mass of the field, and $R$ is the Ricci scalar with a constant $\xi \in \R$. 
A globally hyperbolic spacetime can be shown to be foliated by a one-parameter family of Cauchy surfaces, which leads to the existence of well-defined solutions to the Klein-Gordon equation on $\mfd$ \cite{Wald:QFTCS}.

Let $f \in C_0^\infty(\mfd)$ be a compactly supported real-valued smooth function on $\mfd$. 
In AQFT, a scalar field is instead regarded as an $\R$-valued linear map from $C_0^\infty(\mfd)$ to a unital $*$-algebra\footnote{Here, $*$-algebra $\mathcal{A}$ is an algebra over $\C$ equipped with an involution $*:\mathcal{A}\to \mathcal{A}$ obeying (i) anti-linearity: $\forall A, B \in \mathcal{A}$ and $\forall \lambda_1, \lambda_2 \in \C$, $(\lambda_1 A + \lambda_2 B)^*=\bar \lambda_1 A^* + \bar \lambda_2 B^*$, where $\bar\lambda_i$ is the complex conjugate of $\lambda_i$; (ii) $\forall A, B \in \mathcal{A}$, $(AB)^*=B^* A^*$; (iii) $\forall A \in \mathcal{A}$, $(A^*)^*=A$. 
In particular, a $*$-algebra is called unital if a multiplicative identity $\id \in \mathcal{A}$ is defined.} known as the canonical commutation relation (CCR) algebra: 
\begin{align}
    \phi: C_0^\infty(\mfd) \to \mathcal{A}(\mfd)\,,
    \quad 
    f\mapsto 
    \phi(f)\,.
\end{align}
The CCR algebra $\mathcal{A}(\mfd)$ is generated by smeared field $\phi(f)$ obeying the following relations \cite{khavkine2015algebraic}, which we refer to as the Klein-Gordon relations: 
\begin{itemize}
    \item \textbf{Linearity}: $\phi(af+bf')=a \phi(f)+ b\phi(f')$ for all $a,b \in \R$ and $f, f' \in C_0^\infty (\mfd)$;
    \item \textbf{Hermiticity}: $\phi(f)^*=\phi(f)$ for all $f\in C_0^\infty (\mfd)$;
    \item \textbf{Klein-Gordon equation}: $\phi(Pf)=0$ for all $f\in C_0^\infty (\mfd)$; 
    \item \textbf{Commutation relation}: $[\phi(f), \phi(f')]=\ii E(f, f') \id$ for all $f, f'\in C_0^\infty (\mfd)$;
\end{itemize}
Here, $E(f,f')$ is known as the causal propagator, which can be defined as follows. 
Let $E^+$ and $E^-$ be the retarded and advanced Green operators on $C_0^\infty (\mfd)$ associated with the Klein-Gordon equation, i.e., $E^\pm : C_0^\infty (\mfd) \to C^\infty(\mfd)$ that solve the inhomogeneous Klein-Gordon equation: $P(E^\pm f)=f$. 
The causal propagator operator is the advanced-minus-retarded operator $E\coloneqq E^- - E^+$ on $C_0^\infty(\mfd)$, which solves the Klein-Gordon equation: 
\begin{align}
    E : C_0^\infty (\mfd) \to \sol_\R(\mfd)\,,
    \quad
    P(Ef)=0\,,
\end{align}
where $\sol_\R(\mfd)$ is a vector space of real solutions to the Klein-Gordon equation. 
The causal propagator $E(f, f') \in \R$ can be understood as a smeared real solutions: 
\begin{align}
    E(f,f')
    &\coloneqq
        \int_\mfd \dvol{g}\,f(\sx)(Ef')(\sx)
    \quad f, f' \in C_0^\infty (\mfd)\,, \label{eq:smeared causal propagator}
\end{align}
where $\dvol{g}\equiv \sqrt{-\det g}\, \dd^{n+1} \sx$ is a volume element.

We note that the field $\phi(f)$ is not an operator that acts on a Hilbert space, but rather it is an abstract entity. 
In fact, the aforementioned definition of $\phi(f)$ is an abstraction of the smeared field operator in QFTCS: 
\begin{align}
    \hat \phi(f)
    &=
        \int_\mfd \dvol{g}\, f(\sx) \hat \phi(\sx)\,.
\end{align}
We will see that $\phi(f)$ can be related to the operator $\hat \phi(f)$ on a Hilbert space when a representation is introduced.

We now define an \textit{algebraic state} $\omega$, which corresponds to an expectation value of an observable. 
Formally, the algebraic state is defined as a linear functional $\omega : \mathcal{A}(\mfd) \to \C$ satisfying (i) the normalization condition: $\omega (\id)=1$; and (ii) the positivity condition: $\forall A\in \mathcal{A}(\mfd)$, $\omega (A^* A) \geq 0$. 
We will write a set of algebraic states as $\mathcal{A}^\star(\mfd)$. 
For any two distinct algebraic states $\omega_1$ and $\omega_2$ ($\omega_1 \neq \omega_2$), one can form a new state $\omega$ as a convex linear combination of them: $\omega=\lambda \omega_1 + (1-\lambda)\omega_2$ for $\lambda \in (0,1)$. 
If this is the case, $\omega$ is called \textit{mixed}. 
Otherwise, if an algebraic state $\omega$ cannot be written in this form (thus it is an extreme point in $\mathcal{A}^\star (\mfd)$) then it is called \textit{pure}. 

The algebraic state $\omega$ is related to the states in Hilbert space. 
To see this, we need to introduce a concept of representation of an algebra.

Let us define a \textit{$*$-representation} of $\mathcal{A}(\mfd)$ as a triple $(\mathcal{H}, \mathcal{D}, \pi)$, where $\mathcal{H}$ is Hilbert space, $\mathcal{D}\subset \mathcal{H}$ is a dense subspace of $\mathcal{H}$, and $\pi$ (also referred to as a representation) is a linear homomorphism, 
\begin{align}
    \pi: \mathcal{A}(\mfd) \to \mathcal{L}(\mathcal{D})\,,
    \quad 
    A \mapsto \pi(A)\,,
\end{align}
where $\mathcal{L}(\mathcal{D})$ is the linear space of linear operators that acts on $\mathcal{D}$. 
Here, an involution $*$ is related to a Hermitian adjoint operation in $\mathcal{L}(\mathcal{D})$ in such a way that 
\begin{align}
    \pi(A)^\dag|_\mathcal{D} 
    = 
        \pi(A^*)
    \quad
    \forall A \in \mathcal{A}(\mfd)\,.
\end{align}
For a unital $*$-algebra, we have $\pi(\id)=\id|_{\mathcal{D}}$. 
The field operator on a Hilbert space is then expressed as $\hat \phi(f) \equiv \pi(\phi(f))$.

Given a representation $(\mathcal{H}, \mathcal{D}, \pi)$, each state vector $\ket{\psi} \in \mathcal{H}$ can be related to $\omega$ in such a way that \cite{Hollands.and.Wald2015}
\begin{align}
    \omega_\psi(A)
    &=
        \dfrac{ \braket{\psi|\pi(A)|\psi} }{ \braket{\psi|\psi} }\,, \label{eq:algebraic state and Hilbert space correspondence}
\end{align}
which can also be applied to $\omega_\rho(A)=\Tr[ \hat \rho \pi(A) ]$ for a given density matrix $\hat \rho$. 
A collection of such algebraic states associated to each $\hat \rho$, $S_\pi(\mathcal{A}(\mfd)) \equiv \{ \omega_\rho \} \subset \mathcal{A}^\star(\mfd)$, for a given representation $\pi$ is called a \textit{folium} of $\pi$ \cite{hollands2017entanglement},\footnote{Other literatures often use the terminology `folium of $\omega$', which is equivalent to `folium of $\pi_\omega$ in the GNS representation of $\omega$' in our case.} and each element $\omega_\rho \in S_\pi(\mathcal{A}(\mfd))$ is referred to as a normal (algebraic) state in the representation $\pi$.

Conversely, for each algebraic state $\omega$, there exists a corresponding unique (up to unitary transformation) quadruple $(\mathcal{H}_\omega, \mathcal{D}_\omega, \pi_\omega, \ket{\Omega_\omega})$ satisfying 
\begin{align}
    \omega(A)
    &=
        \braket{\Omega_\omega|\pi_\omega(A)|\Omega_\omega}
    \quad \forall A \in \mathcal{A}(\mfd)\,,
\end{align}
as well as $\pi_\omega(\mathcal{A}(\mfd)) \ket{\Omega_\omega}=\mathcal{D}_\omega$. 
The formation of $(\mathcal{H}_\omega, \mathcal{D}_\omega, \pi_\omega, \ket{\Omega_\omega})$ is known as the Gelfand-Naimark-Segal (GNS) construction.

This correspondence between the algebraic states and the state vectors defined in a Hilbert space tells us that AQFT and the traditional Hilbert space approach of QFTCS are essentially equivalent. 
However, as we previously mentioned, AQFT differs in that it does not necessitate specifying a Hilbert space, or equivalently, selecting a representation $\pi$, until the last step.

So far, we have introduced a CCR algebra and its $*$-representation. 
However, the operators $\pi(A) \in \mathcal{L}(\mathcal{D})$ are generally unbounded, which can lead to certain challenges such as domain issues. 
To overcome such challenges, we use an exponentiated version of $\mathcal{A}(\mfd)$ called the \textit{Weyl algebra}.

The Weyl algebra $\mathcal{W}(\mfd)$ is a unital $C^*$-algebra generated by $W(Ef) \equiv e^{ \ii \phi(f) }$ obeying the \textit{Weyl relations}: 
\begin{subequations}
    \begin{align}
        &W(Ef)^*
        =
            W(-Ef)\,,\\
        &W(Ef_1) W(Ef_2)
        =
            e^{ -\ii E(f_1, f_2)/2 } W(E(f_1 + f_2))\,.
    \end{align} \label{eq:Weyl relations}
\end{subequations}
We note that $W(0)=\id$ is the identity element of $\mathcal{W}(\mfd)$. 
Then, the notion of the algebraic state $\omega$ carries over as $\omega : \mathcal{W}(\mfd) \to \C$ (we denote the linear space of $\omega$ by $\mathcal{W}^\star(\mfd)$). 
Moreover, a representation of $C^*$-algebra provides a set of bounded linear operators $\mathcal{B}(\mathcal{H})$ on a Hilbert space $\mathcal{H}$. 

\subsection{Quasi-free states}\label{sec:quasifree states}
The \textit{Wightman $n$-point function} is an essential quantity in QFT. 
For $f_j \in C_0^\infty (\mfd)$ and $\phi(f_j) \in \mathcal{A}(\mfd)$, the Wightman $n$-point function in an algebraic state $\omega$ is defined as 
\begin{align}
    \mathsf{W}(f_1, f_2, ..., f_n)
    \coloneqq
        \omega (\phi(f_1) \phi(f_2)...\phi(f_n))\,.
\end{align}

We now restrict ourselves to a class of physically admissible algebraic states known as Hadamard states \cite{Hollands.and.Wald2015, KayWald1991theorems}. 
Hadamard algebraic states are characterized by their Wightman $n$-point functions, which, at short distances, exhibit behaviors similar to those in the Minkowski vacuum state in Minkowski spacetime. 
A subset of these Hadamard states allows for the expression of $n$-point functions using only one- and two-point Wightman functions, which we refer to as Gaussian states. 
Of particular interest are the \textit{quasi-free states}, a subset of Gaussian states, where the one-point Wightman functions vanish: $\omega(\phi(f))=0$.\footnote{In the literature, quasi-free states are also called Gaussian states. 
Our use of 'quasi-free' specifically refers to 'quasi-free states with vanishing one-point Wightman function.' \cite{KayWald1991theorems}} 
In these quasi-free algebraic states, every even-point Wightman function can be expressed only in terms of the two-point Wightman functions $\omega(\phi(f_1) \phi(f_2))$, while all odd-point functions vanish. 
For example, while a coherent state is a non-quasi-free Gaussian state, states like vacuum, squeezed, and thermal Kubo-Martin-Schwinger (KMS) states fall under the quasi-free type.

The properties above also hold for the Weyl algebra $\mathcal{W}(\mfd)$. 
In this case, the two-point Wightman function is understood as \cite{KayWald1991theorems}
\begin{align}
    \mathsf{W}(f_1, f_2)
    &=
        -\dfrac{ \partial^2 }{ \partial s \partial t }
        \bigg|_{s,t=0}
        \omega( W(sEf_1) W(tEf_2) )\,,
\end{align}
and if $\omega: \mathcal{W}(\mfd) \to \C$ is a quasi-free state, then we have 
\begin{align}
    \omega(W(Ef))
    &=
        e^{ -\frac12 \mathsf{W}(f,f) }\,. \label{eq:quasifree state}
\end{align}
In what follows, we will first consider a general algebraic state $\omega$ and then impose \eqref{eq:quasifree state} at the end to classify $\omega$ as a quasi-free state. 
It will be useful to decompose the two-point Wightman function into its real and imaginary parts as 
\begin{align}
    \mathsf{W}(f_1, f_2)
    &=
        \text{Re}[ \mathsf{W}(f_1, f_2) ]
        + \ii\, \text{Im}[\mathsf{W}(f_1, f_2)] 
    \equiv 
        \mu (Ef_1, Ef_2) 
        + \dfrac{\ii }{2} E(f_1, f_2)\,, \label{eq:wightman and mu and causal propagator}
\end{align}
where $\mu(Ef_1, Ef_2) \in \R$ originates from an inner product of solutions to the Klein-Gordon equation (see \cite{Wald:QFTCS, KayWald1991theorems} for details), and $E(f_1, f_2) \in \R$ is the causal propagator given in \eqref{eq:smeared causal propagator}.

\section{Unruh-DeWitt Detector Model}\label{sec: UDW}

We now introduce our qubit model in the quantum Otto cycle. 
The qubit model we employ here is called the Unruh-DeWitt (UDW) particle detector model \cite{Unruh1979evaporation, DeWitt1979}. 
A UDW detector has ground $\ket{g}$ and excited $\ket{e}$ states and their energy gap is denoted by $\Omega$. 
The free Hamiltonian $\hat H\ts{D,0}$ is given by 
\begin{align}
    \hat H\ts{D,0}
    &=
        \dfrac{\Omega}{2} ( \hat \sigma_z + \id\ts{D} )\,,
\end{align}
where $\hat \sigma_z$ is the Pauli-$z$ operator and $\id\ts{D}$ is the identity operator on the detector's Hilbert space.

We then let a detector interact with the field $\hat \phi$ in a spacetime region specified by $f\in C_0^\infty(\mfd)$ in the following manner. 
In the interaction picture, the time-evolution unitary operator $\hat U\ts{I}$ is given by 
\begin{align}
    \hat U\ts{I}
    &=
        \mathcal{T}_\tau 
        \exp
        \kako{
            -\ii 
            \int_\mfd 
            \dvol{g}\, \hat h\ts{I}(\sx)
        }\,,
    \quad
        \hat h\ts{I}(\sx)
        \coloneqq
            f(\sx) \hat \mu(\tau(\sx)) \otimes \hat \phi(\sx)
    \label{eq:time-ordered unitary}
\end{align}
where $\mathcal{T}_\tau$ is a time-ordering symbol with respect to the proper time $\tau$ of the detector, and $\hat \mu(\tau(\sx))$ is a monopole moment operator given by 
\begin{align}
    \hat \mu(\tau(\sx))
    &=
        e^{\ii \Omega \tau(\sx)} \ket{e} \bra{g}
        +
        e^{-\ii \Omega \tau(\sx)} \ket{g} \bra{e}\,.
\end{align}

As mentioned earlier, $f\in C_0^\infty(\mfd)$ is the interaction region between the detector and the field. 
Roughly speaking, one can think of the spatial extension of $f$ on each time slice as the `shape' of the detector, and its timelike extension as the `interaction duration.' 
In fact, it is possible to separate these parts in $f$ by adopting a suitable coordinate system called the \textit{Fermi normal coordinates} $\bar \sx = (\tau, \bar \bx)$, which is adapted to the trajectory of the detector's center-of-mass. 
In this coordinate system, $\bar \bx= \bm{0}$ is defined to be the position of the detector's center-of-mass, and $\tau$ is its proper time. 
Then the interaction region $f$ can be decomposed into 
\begin{align}
    f(\sx(\bar \sx))
    &=
        \lambda \chi(\tau/\eta) F(\bar\bx)\,,
\end{align}
where $\lambda \geq 0$ is the coupling constant, $\chi(\tau/\eta)$ is a switching function (with $\eta>0$ a quantity in units of time) describing the time-dependence of interaction, and $F(\bar \bx)$ is a smearing function representing the shape of the detector.

It is well-known that the existence of the time-ordering symbol in \eqref{eq:time-ordered unitary} prevents us from obtaining an exact solution. 
As a result, one might consider employing a perturbative analysis for further exploration. 
In this paper, however, we will specifically focus on a special class of $\hat h\ts{I}$ known as \textit{delta-coupling} \cite{Simidzija.Nonperturbative, Funai.negative.energy.delta, Simidzija2018no-go, Henderson.Bandlimited, SahuSabotage, GallockEntangledDetectors, Perche.curvature.delta, Tjoa.RQC, Shalabi.locally.UV.delta, Tjoa.nonperturbative.gaussian, Jose.tran.delta} to carry out a non-perturbative analysis. 
The delta-coupling approach is characterized by an interaction time scale that is significantly shorter than the typical time scale relevant to the system under study. 
The switching function in the Fermi normal coordinates in the delta-coupling approach is given by 
\begin{align}
    \chi((\tau - \tau_0)/\eta)
    &=
        \eta \delta(\tau-\tau_0)\,,
\end{align}
where $\delta$ is the Dirac delta distribution, $\tau_0 \in \R$ is the time when the detector switches on and off instantaneously, and we used the identity $\delta(\tau/\eta)=\eta \delta(\tau)$. 
Since it only deals with an instant time, the time-ordering will not contribute to the unitary operator in \eqref{eq:time-ordered unitary}, thereby the resulting $\hat U\ts{I}$ reads \cite{Jose.tran.delta}
\begin{align}
    \hat U\ts{I}
    &=
        \exp 
        \kagikako{
            -\ii \hat \mu(\tau_0) \otimes \hat \phi(\mathfrak{f})
        }\,, \label{eq:Schmidt rank1 unitary}
\end{align}
where
\begin{align}
    \hat \phi(\mathfrak{f})
    \coloneqq
        \int_{\Sigma_{\tau_0}} \dd \Sigma\, 
        \lambda \eta F(\bar \bx) \hat \phi(\tau_0, \bar\bx) \label{eq:spatially smeared field}
\end{align}
with $\dd \Sigma$ being a volume element on a time slice $\Sigma_{\tau_0}$ at $\tau=\tau_0$. 
The function $\mathfrak{f} \in C_0^\infty (\Sigma_{\tau_0})$ can be thought of as a spatial version of $f\in C_0^\infty(\mfd)$. 
The unitary operator of the form \eqref{eq:Schmidt rank1 unitary} is a particular example of the so-called Schmidt rank-1 operators.\footnote{A Schmidt rank-$r$ unitary operator $\hat U$ on a bipartite system $\mathcal{H}\ts{A}\otimes \mathcal{H}\ts{B}$ is defined as $\hat U= \sum_{i=1}^r c_i \hat A_i \otimes \hat B_i$ with $\forall c_i \neq 0$, where $\{ \hat A_i \}$ and $\{ \hat B_i \}$ are linearly independent operators.} 
Using a property $\hat \mu(\tau_0)^2 = \id\ts{D}$, Eq.~\eqref{eq:Schmidt rank1 unitary} can be expressed as 
\begin{align}
    \hat U\ts{I}
    &=
        \mathfrak{\id}\ts{D} 
        \otimes \cos (\hat \phi(\mathfrak{f}))
        - 
        \ii \hat \mu(\tau_0) \otimes \sin(\hat \phi(\mathfrak{f}))\,.
\end{align}

\begin{figure}[t]
\centering
\includegraphics[width=0.8\linewidth]{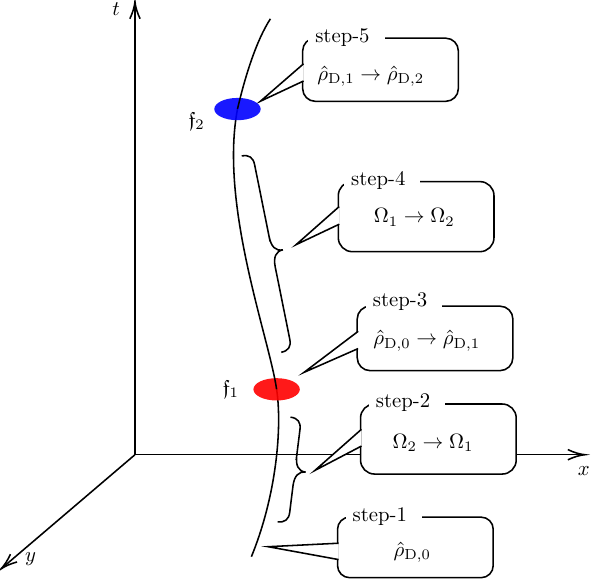}
\caption{A UDW detector interacting with the field instantaneously.}
\label{fig:deltaQOE}
\end{figure}

In the quantum Otto cycle, the detector interacts with the field twice, as depicted in Figure~\ref{fig:deltaQOE}. 
In this scenario, the unitary operator becomes 
\begin{align}
    \hat U\ts{I}
    &=
        \mathcal{T}_\tau 
        \exp
        \kako{
            -\ii 
            \int_\mfd 
            \dvol{g}\, 
            \kagikako{
                \hat h\ts{I,1}(\sx) + \hat h\ts{I,2}(\sx)
            }
        }\,,\label{eq:time-ordered unitary twice}
\end{align}
where $\hat h\ts{I,$j$}(\sx)$ $j\in \{ 1,2 \}$ corresponds to the first ($j=1$) and the second ($j=2$) interaction between the field and the detector with energy gap $\Omega_j$ in the region $f_j \in C_0^\infty(\mfd)$, respectively. 
Employing the delta-coupling for each interaction, Eq.~\eqref{eq:time-ordered unitary twice} can be written as a product of Schmidt rank-1 operators \cite{Jose.tran.delta}: 
\begin{align}
    \hat U\ts{I}
    &=
        \hat U\ts{I}^{(2)} \hat U\ts{I}^{(1)}\,,\label{eq:universal time evolution} \\
    \hat U\ts{I}^{(j)}
    &\coloneqq
        \exp
        \kagikako{
            -\ii \hat \mu(\tau_{j}) \otimes \hat \phi(\mathfrak{f}_j)
        }
    =
        \mathfrak{\id}\ts{D}\otimes \cos (\hat \phi(\mathfrak{f}_j))
        - 
        \ii \hat \mu(\tau_j) \otimes \sin(\hat \phi(\mathfrak{f}_j))
    \,.\label{eq:unitary in terms of C and S}
\end{align}
Here, we assumed $\tau_{2} > \tau_1$ (i.e., the first interaction at $\tau=\tau_1$ is in the causal past of the second interaction at $\tau=\tau_2$). 
In the following discussion, we will write $\hat \mu_j \equiv \hat \mu(\tau_j)$ and $\hat \phi_j \equiv \hat \phi(\mathfrak{f}_j)$ for simplicity.

\section{Results: Extracted Work from QOE}\label{sec: results}

We now consider a single UDW detector operating within the quantum Otto cycle. 
The background is a globally hyperbolic $(n+1)$-dimensional spacetime, in which we define a quantum Klein-Gordon field in a quasi-free state. 
The interaction between the detector and the field is described by the Schmidt rank-1 unitary operator \eqref{eq:universal time evolution}. 
To maintain generality, we begin our calculations with a generic algebraic state $\omega$, then later specify it to a quasi-free state.

\subsection{General formula of work} \label{subsec:general formula of work}

\subsubsection{The first interaction: from step-1 to step-3}\label{subsec:first interaction}

As step-\ref{step:step1} in the quantum Otto cycle, let the initial state of the total system be 
\begin{align}
    \hat \rho\ts{tot,0}
    =
        \hat \rho\ts{D,0}\otimes \hat \rho_{\phi,0}\,,
\end{align}
where $\hat \rho\ts{D,0}$ and $\hat \rho_{\phi,0}$ are the initial state of the detector and the field, respectively, and $\hat \rho\ts{D,0}$ is specified to be 
\begin{align}
    \hat \rho\ts{D,0}
    &=
        p \ket{e}\bra{e} + (1-p) \ket{g}\bra{g}\,,
        \quad p\in [0,1]\,.\label{eq:initial state}
\end{align}
As depicted in Figure~\ref{fig:deltaQOE}, the initial energy gap is $\Omega_2$ and then adiabatically expanded to $\Omega_1(>\Omega_2)$ in step-\ref{step:step2}.

In step-\ref{step:step3} (the isochoric process) of the cycle, the detector interacts with the field within the region $\mathfrak{f}_1$, where the detector's proper time is $\tau=\tau_1$. 
The resulting state of the total system is 
\begin{align}
    \hat \rho\ts{tot,1}
    &=
        \hat U\ts{I}^{(1)} \hat \rho\ts{tot,0} \hat U\ts{I}^{(1)\dagger} \notag \\
    &=
        \hat \rho\ts{D,0}
        \otimes \hat C_1 \hat \rho_{\phi,0} \hat C_1
        +
        \hat \mu_1 \hat \rho\ts{D,0} \hat \mu_1
        \otimes \hat S_1 \hat \rho_{\phi,0} \hat S_1
        -
        \ii \hat \mu_1 \hat \rho\ts{D,0}
        \otimes 
        \hat S_1 \hat \rho_{\phi,0} \hat C_1
        +
        \ii \hat \rho\ts{D,0} \hat \mu_1
        \otimes 
        \hat C_1 \hat \rho_{\phi,0} \hat S_1\,,
\end{align}
where we have used \eqref{eq:unitary in terms of C and S} and wrote $\hat S_1\equiv \sin(\hat \phi_1)$ and $\hat C_1 \equiv \cos (\hat \phi_1)$ for simplicity. 
We note that $[\hat C_1, \hat S_1]=0$ is used. 
The density matrix of the detector after the first interaction is obtained by tracing out the field degree of freedom:  
\begin{align}
    \hat \rho\ts{D,1}
    &=
        \Tr_\phi[ \hat U\ts{I}^{(1)} \hat \rho\ts{tot,0} \hat U\ts{I}^{(1)\dagger} ] \notag \\
    &=
        \hat \rho\ts{D,0}
        \omega(C_1^2)
        +
        \hat \mu_1 \hat \rho\ts{D,0} \hat \mu_1
        \omega(S_1^2)
        +
        \ii [\hat \rho\ts{D,0}, \hat \mu_1]
        \omega(S_1 C_1)\,, \label{eq:rhoD1 with algebraic states}
\end{align}
where $C_1$ and $S_1$ are the elements of the Weyl algebra corresponding to $\hat C_1$ and $\hat S_1$, respectively, via $\pi(C_1)\equiv \hat C_1$ and $\pi(S_1)\equiv \hat S_1$ in a representation $\pi$ that gives $\hat \rho_{\phi,0}$ in its folium. 
In this notation, $\omega (A)=\Tr[ \hat \rho_{\phi,0} \hat A ]$ is used.

We now calculate each algebraic state in \eqref{eq:rhoD1 with algebraic states}. 
Let us write $C_1$ and $S_1$ in terms of the elements of the Weyl algebra $W_{\mathfrak{f}_j} \equiv W(E\mathfrak{f}_j)$: 
\begin{subequations}
    \begin{align}
        &C_1
        =
            \dfrac{ e^{\ii \phi_1} + e^{-\ii \phi_1} }{2}
        =
            \dfrac{ W_{\mathfrak{f}_1} + W_{\mathfrak{f}_1}^* }{2}\,, \\
        &S_1
        =
            \dfrac{ e^{\ii \phi_1} - e^{-\ii \phi_1} }{2\ii}
        =
            \dfrac{ W_{\mathfrak{f}_1} - W_{\mathfrak{f}_1}^* }{2\ii}\,.
    \end{align}\label{eq:S and C in terms of Weyl}
\end{subequations}
These can be written in a unified manner as 
\begin{align}
    \dfrac{ W_{\mathfrak{f}_1} +(-1)^u W_{\mathfrak{f}_1}^* }{2\ii^{u}}\,,
    \quad
    u \in \{ 0,1 \}
\end{align}
where $u=0$ and $u=1$ correspond to $C_1$ and $S_1$, respectively. 
Then each algebraic state in \eqref{eq:rhoD1 with algebraic states} takes the form 
\begin{align}
    &\omega 
    \kako{
        \dfrac{ W_{\mathfrak{f}_1} +(-1)^r W_{\mathfrak{f}_1}^* }{2\ii^r}
        \dfrac{ W_{\mathfrak{f}_1} +(-1)^s W_{\mathfrak{f}_1}^* }{2\ii^s}
    } \,.
\end{align}
Using the Weyl relations \eqref{eq:Weyl relations} and the fact that $\omega: \mathcal{W}(\mathcal{M}) \to \mathbb{C}$ is a linear map, we get
\begin{align}
    &\omega 
    \kako{
        \dfrac{ W_{\mathfrak{f}_1} +(-1)^r W_{\mathfrak{f}_1}^* }{2\ii^r}
        \dfrac{ W_{\mathfrak{f}_1} +(-1)^s W_{\mathfrak{f}_1}^* }{2\ii^s}
    } \notag \\
    &=
        \dfrac{1}{4\ii^{r+s}}
        \Bigkagikako{
            \omega(W_{2\mathfrak{f}_1})
            +
            (-1)^r
            +
            (-1)^s
            +
            (-1)^{r+s} \omega(W_{2\mathfrak{f}_1}^* )
        }\,.
\end{align}

So far, we have assumed that the state of the field is arbitrary. 
We now specify $\hat \rho_{\phi,0}$ to be a quasi-free state defined in \eqref{eq:quasifree state}. 
We then define 
\begin{align}
    \nu_{\mathfrak{f}_1} 
    \equiv
        \omega(W_{2\mathfrak{f}_1})
    =
        e^{-2\mathsf{W}(\mathfrak{f}_1, \mathfrak{f}_1)} \in (0,1]\,. \label{eq:nu_f}
\end{align}
In this case, the algebraic state is a real number, which leads to $\omega(W_{\mathfrak{f}}^*)=\omega(W_{\mathfrak{f}})$. 
Then 
\begin{align}
    \omega 
    \kako{
        \dfrac{ W_{\mathfrak{f}_1} +(-1)^r W_{\mathfrak{f}_1}^* }{2\ii^r}
        \dfrac{ W_{\mathfrak{f}_1} +(-1)^s W_{\mathfrak{f}_1}^* }{2\ii^s}
    } 
    &=
        \dfrac{1}{4\ii^{r+s}}
        \Bigkagikako{
            (1 + (-1)^{r+s}) \nu_{\mathfrak{f}_1}
            +
            (-1)^r
            +
            (-1)^s
        }\,.
\end{align}
This tells us that $\omega(S_1 C_1)=\omega(C_1 S_1)=0$ and so the detector's state after the first interaction reads
\begin{align}
    \hat \rho\ts{D,1}
    &=
        \hat \rho\ts{D,0}
        \omega(C_1^2)
        +
        \hat \mu_1 \hat \rho\ts{D,0} \hat \mu_1
        \omega(S_1^2) \notag \\
    &=
        \dfrac{ 1+\nu_{\mathfrak{f}_1} }{ 2 }
        \hat \rho\ts{D,0}
        +
        \hat \mu_1 \hat \rho\ts{D,0} \hat \mu_1
        \dfrac{ 1 - \nu_{\mathfrak{f}_1} }{ 2 } \\
    &\equiv
        p_1 \ket{e}\bra{e} + (1-p_1) \ket{g}\bra{g}\,,
\end{align}
where 
\begin{align}
    p_1
    &=
        \dfrac{1}{2} 
        + 
        \kako{
            p-\dfrac{1}{2}
        } 
        \nu_{\mathfrak{f}_1}\,, \label{eq:p1}
\end{align}
when $\hat \rho\ts{D,0}$ is initially chosen to be \eqref{eq:initial state}.

\subsubsection{The second interaction: from step-4 to step-5}

Next, we move to the adiabatic compression process $\Omega_1 \to \Omega_2$ in step-\ref{step:step4}, which is then followed by a second interaction within the region $\mathfrak{f}_2$ during step-\ref{step:step5}.

The final state of the total system, $\hat \rho\ts{tot,2}$, is given by 
\begin{align}
    \hat \rho\ts{tot,2}
    &=
        \hat U\ts{I}^{(2)}
        \hat \rho\ts{tot,1}
        \hat U\ts{I}^{(2)\dagger}
    =
        \hat U\ts{I}^{(2)} \hat U\ts{I}^{(1)}
        \hat \rho\ts{tot,0}
        \hat U\ts{I}^{(1)\dagger} \hat U\ts{I}^{(2)\dagger}\,,
\end{align}
where $\hat U\ts{I}^{(j)}$ is given by \eqref{eq:unitary in terms of C and S}. 
We will again write $\hat S_j\equiv \sin(\hat \phi_j)$ and $\hat C_j \equiv \cos (\hat \phi_j)$ and use the properties $[\hat C_j, \hat S_j]=0$ and $[\hat C_j, \hat S_k] \neq 0$ for $j\neq k$.

After the second interaction in step-\ref{step:step5}, the final density matrix for the detector, $\hat \rho\ts{D,2}$, can be straightforwardly derived as in the previous case (see Appendix~\ref{app:second interaction calculation} for the detailed calculations): 
\begin{align}
    \hat \rho\ts{D,2}
    &=
        \hat \rho\ts{D,0}  \omega(C_1 C_2^2 C_1) 
        + 
        \hat \mu_2 \hat \rho\ts{D,0} \hat \mu_2 \omega(C_1 S_2^2 C_1) 
        +
        \hat \mu_1 \hat \rho\ts{D,0} \hat \mu_1 \omega(S_1 C_2^2 S_1) \notag \\
        &\quad
        +
        \hat \mu_2 \hat \mu_1 \hat \rho\ts{D,0} \hat \mu_1 \hat \mu_2 \omega(S_1 S_2^2 S_1) 
        +
        [\hat \mu_1 \hat \rho\ts{D,0}, \hat \mu_2] \omega(C_1 S_2 C_2 S_1) 
        - 
        [\hat \rho\ts{D,0}\hat \mu_1, \hat \mu_2] \omega(S_1 S_2 C_2 C_1)\,,
\end{align}
where 
\begin{subequations}
    \begin{align}
        &\omega(C_1 C_2^2 C_1)
        =
            \dfrac{1}{4}
            \Bigkagikako{
                1 
                + \nu_{\mathfrak{f}_1}
                + \nu_{\mathfrak{f}_1} \nu_{\mathfrak{f}_2} \cosh[ 4\mu(E\mathfrak{f}_1, E\mathfrak{f}_2) ]
                + \nu_{\mathfrak{f}_2} \cos (2E(\mathfrak{f}_1, \mathfrak{f}_2)) 
            }\,,\\
        &\omega(C_1 S_2^2 C_1)
        =
            \dfrac{1}{4}
            \Bigkagikako{
                1 
                + \nu_{\mathfrak{f}_1}
                - \nu_{\mathfrak{f}_1} \nu_{\mathfrak{f}_2} \cosh[ 4\mu(E\mathfrak{f}_1, E\mathfrak{f}_2) ]
                - \nu_{\mathfrak{f}_2} \cos (2E(\mathfrak{f}_1, \mathfrak{f}_2)) 
            }\,,\\
        &\omega(S_1 C_2^2 S_1)
        =
            \dfrac{1}{4}
            \Bigkagikako{
                1 
                - \nu_{\mathfrak{f}_1}
                - \nu_{\mathfrak{f}_1} \nu_{\mathfrak{f}_2} \cosh[ 4\mu(E\mathfrak{f}_1, E\mathfrak{f}_2) ]
                + \nu_{\mathfrak{f}_2} \cos (2E(\mathfrak{f}_1, \mathfrak{f}_2)) 
            }\,,\\
        &\omega(S_1 S_2^2 S_1)
        =
            \dfrac{1}{4}
            \Bigkagikako{
                1 
                - \nu_{\mathfrak{f}_1}
                + \nu_{\mathfrak{f}_1} \nu_{\mathfrak{f}_2} \cosh[ 4\mu(E\mathfrak{f}_1, E\mathfrak{f}_2) ]
                - \nu_{\mathfrak{f}_2} \cos (2E(\mathfrak{f}_1, \mathfrak{f}_2)) 
            }\,,\\
        &\omega(C_1 C_2 S_2 S_1)
        =
            \omega(C_1 S_2 C_2 S_1)
        =
            \dfrac{\nu_{\mathfrak{f}_2}}{4}
            \Bigkagikako{
                \nu_{\mathfrak{f}_1} \sinh[ 4\mu(E\mathfrak{f}_1, E\mathfrak{f}_2) ]
                - \ii \sin (2E(\mathfrak{f}_1, \mathfrak{f}_2)) 
            }\,,\\
        &\omega(S_1 C_2 S_2 C_1)
        =
            \omega(S_1 S_2 C_2 C_1)
        =
            \dfrac{\nu_{\mathfrak{f}_2}}{4}
            \Bigkagikako{
                \nu_{\mathfrak{f}_1} \sinh[4 \mu(E\mathfrak{f}_1, E\mathfrak{f}_2) ]
                + \ii \sin (2E(\mathfrak{f}_1, \mathfrak{f}_2)) 
            }\,.
    \end{align}
\end{subequations}
Here, $\mu(E\mathfrak{f}_1, E\mathfrak{f}_2) \in \R$ is given in \eqref{eq:wightman and mu and causal propagator}. 
Inserting these expressions, the final density matrix $\hat \rho\ts{D,2}$ reads 
\begin{align}
    \hat \rho\ts{D,2}
    &=
        p_2 \ket{e}\bra{e} + (1-p_2) \ket{g}\bra{g}\,,
\end{align}
where 
\begin{align}
    p_2
    &=
        \dfrac{1}{2}
        \kagikako{
            1
            + \nu_{\mathfrak{f}_2} \sin(2E(\mathfrak{f}_1, \mathfrak{f}_2)) \sin \theta 
            + (2p-1) \nu_{\mathfrak{f}_1} \nu_{\mathfrak{f}_2}
            \kako{
                e^{4 \mu(E\mathfrak{f}_1, E\mathfrak{f}_2)}
                \sin^2\dfrac{\theta}{2}
                +
                e^{-4 \mu(E\mathfrak{f}_1, E\mathfrak{f}_2)}
                \cos^2\dfrac{\theta}{2}
            }
        }\,, \label{eq:p2} 
\end{align}
where $\theta \coloneqq \Omega_1 \tau_1 - \Omega_2 \tau_2$. 
Note that 
\begin{align}
    0 < \nu_{\mathfrak{f}_1} \nu_{\mathfrak{f}_2}
            \kako{
                e^{4 \mu(E\mathfrak{f}_1, E\mathfrak{f}_2)}
                \sin^2\dfrac{\theta}{2}
                +
                e^{-4 \mu(E\mathfrak{f}_1, E\mathfrak{f}_2)}
                \cos^2\dfrac{\theta}{2}
            }
        \leq 1\,. \label{eq:pee pee poo poo}
\end{align}

\subsubsection{Closing the cycle and work extracted: step-6}
Let us complete the thermodynamic cycle and evaluate the work extracted by the delta-coupled UDW detector.

We first impose the cyclicity condition in step-\ref{step:step6}, $p_2=p$, which can be recast into
\begin{align}
    p
    &=
        \dfrac{1}{2}
        -
        \dfrac{ \dfrac{\nu_{\mathfrak{f}_2}}{2}
        \sin(2E(\mathfrak{f}_1,\mathfrak{f}_2)) 
        \sin \theta }
        {
            \nu_{\mathfrak{f}_1} \nu_{\mathfrak{f}_2} 
            \kako{
                e^{4 \mu(E\mathfrak{f}_1, E\mathfrak{f}_2)}
                \sin^2\dfrac{\theta}{2}
                +
                e^{-4 \mu(E\mathfrak{f}_1, E\mathfrak{f}_2)}
                \cos^2\dfrac{\theta}{2}
            }
            -1
        }\,.\label{eq:cyclicity in terms of p}
\end{align}
This means the initial population $p$ in the excited state $\ket{e}$ has to be adjusted according to \eqref{eq:cyclicity in terms of p} to ensure that the thermodynamic cycle is closed.

Under the cyclicity condition \eqref{eq:cyclicity in terms of p}, let us calculate the work extracted from the field [see \eqref{eq:total extracted work}]. 
From \eqref{eq:p1}, we obtain
\begin{align}
    \braket{W\ts{ext}}
    &=
        (p_1-p)\Delta \Omega \notag \\
    &=
        -
        \kako{
            p-\dfrac{1}{2}
        }
        (1-\nu_{\mathfrak{f}_1}) \Delta \Omega \\
    &=
        \dfrac{
            \dfrac{\nu_{\mathfrak{f}_2}}{2}
            \sin(2E(\mathfrak{f}_1,\mathfrak{f}_2)) 
            \sin\theta
            (1-\nu_{\mathfrak{f}_1}) 
        }
        {
            \nu_{\mathfrak{f}_1} \nu_{\mathfrak{f}_2} 
            \kako{
                e^{4 \mu(E\mathfrak{f}_1, E\mathfrak{f}_2)}
                \sin^2\dfrac{\theta}{2}
                +
                e^{-4 \mu(E\mathfrak{f}_1, E\mathfrak{f}_2)}
                \cos^2\dfrac{\theta}{2}
            }
            -1
        }
        \Delta \Omega\,, \label{eq:work extracted by delta switch}
\end{align}
where $p$ in \eqref{eq:cyclicity in terms of p} is substituted to meet the cyclicity condition. 
Recalling that $\nu_\mathfrak{f} \in (0,1]$, $\Delta \Omega > 0$, and Eq.~\eqref{eq:pee pee poo poo}, the positive work condition, $\braket{W\ts{ext}}>0$, reads 
\begin{align}
    \sin(2E(\mathfrak{f}_1,\mathfrak{f}_2)) 
    \sin \theta < 0 \,. \label{eq:positive work condition for delta}
\end{align}
Therefore, we obtained the work extracted \eqref{eq:work extracted by delta switch} and its positive work condition \eqref{eq:positive work condition for delta} for a UDW detector, which instantaneously interacts with a quantum scalar field in a quasi-free state in an arbitrary globally hyperbolic spacetime.

\subsection{Observations and remarks}\label{subsec: observation}

\begin{figure}[t]
\centering
\includegraphics[width=0.5\linewidth]{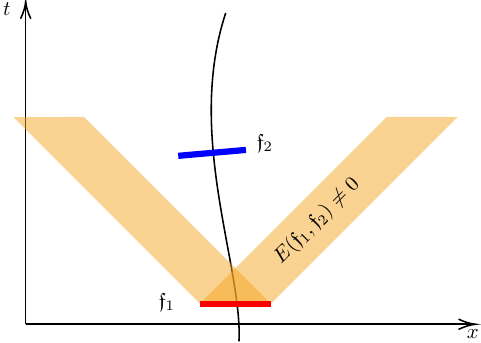}
\caption{Image of self-communication. 
For a massless scalar field in $(3+1)$-dimensional Minkowski spacetime, the field propagates in the lightlike direction. 
The detector can receive a signal from itself in the past if the interaction region $\mathfrak{f}_2$ intersects with the yellow lightlike region. 
Otherwise, $E(\mathfrak{f}_1, \mathfrak{f}_2)=0$.}
\label{fig:communication}
\end{figure}

Let us discuss the implication of the final results \eqref{eq:work extracted by delta switch} and \eqref{eq:positive work condition for delta}. 
\begin{itemize}
\item \textit{Work from signaling;}

We observe from \eqref{eq:positive work condition for delta} that a QOE with instantaneous isochoric processes is capable of extracting positive work, depending on the causal propagator $E(\mathfrak{f}_1, \mathfrak{f}_2)$. 
The causal propagator $E(\mathfrak{f}_1, \mathfrak{f}_2)$ represents the signaling between the regions $\mathfrak{f}_1$ and $\mathfrak{f}_2$ through the quantum field. 
The vanishing causal propagator, $E(\mathfrak{f}_1, \mathfrak{f}_2)=0$, implies no signaling between the regions. 
Since we only have a single UDW detector, this signaling refers to a self-communication, i.e., the detector in region $\mathfrak{f}_2$ receiving its own signal sent from $\mathfrak{f}_1$ when it first coupled to the field. 
This point is visualized in Figure~\ref{fig:communication}.

The positive work condition \eqref{eq:positive work condition for delta} implies that if the detector is unable to self-communicate, then $E(\mathfrak{f}_1, \mathfrak{f}_2)=0$ leading to $\braket{W\ts{ext}}=0$. 
Thus, self-communication is necessary to extract positive work for a delta-coupled UDW detector. 
We stress that this holds for any globally hyperbolic spacetimes, though the way signals propagate may vary depending on the spacetime under consideration. 
It should be noted that merely receiving a signal from the past does not guarantee work extraction since the sign in \eqref{eq:positive work condition for delta} might be positive even when $E(\mathfrak{f}_1,\mathfrak{f}_2) \neq 0$. 
We will address the mechanism for the work extraction at the end of this subsection.

\item \textit{The state of the field;}

The causal propagator $E(\mathfrak{f}_1, \mathfrak{f}_2)$ is a state-independent quantity, i.e., its value is the same no matter what the field state $\omega$ is. 
Therefore, the positive work condition \eqref{eq:positive work condition for delta} is state-independent as well.

\item \textit{Extracting work from a vacuum state and the second law;}

As we will discuss in Section~\ref{subsec:Example of Minkowski}, it is possible to extract positive work even when an inertial detector interacts with the vacuum state $\ket{0}$ in Minkowski spacetime. 
This can be qualitatively explained as follows. 
The second law of thermodynamics states that work cannot be extracted from a system in equilibrium. 
However, if the system is brought out of equilibrium, extracting work becomes feasible. 
A typical example is a system with two thermal baths at different temperatures. 
In our case, the quantum field becomes an out-of-equilibrium state due to the disturbance caused by the first interaction. 
If the detector interacts with this disturbed field, it may be possible to extract work. 


\item \textit{Comparison to a perturbative analysis;}

A perturbative analysis of QOE utilizing a UDW detector was investigated in \cite{Gallock2023Otto} (see \cite{UnruhOttoEngine, Finn.UnruhOtto, Xu.UnruhOtto.degenerate, Kane.entangled.Unruh.Otto, Barman.entangled.UnruhOtto, Mukherjee.UnruhOtto.boundary} for analyses involving a linearly accelerating UDW detector). 
In these scenarios, a pointlike (i.e., zero-size) UDW detector interacts with a quasi-free state over a finite time duration. 
The extracted work depends only on the response function (the probability of transition between $\ket{g} \leftrightarrow \ket{e}$) and communication is irrelevant in the leading order of perturbation theory. 
See also \cite{Causality2015Eduardo} for the communication effect involving two detectors.

Based on the findings presented in this paper, we expect that a weakly-coupled UDW detector may be able to extract work from a vacuum state if higher-order terms are accounted for in the perturbation theory. 
\end{itemize}

Before proceeding to the next subsection, let us examine the role of signaling in our QOE. 
To understand the work extraction mechanism, let us remind ourselves the condition for positive work extraction from QOE. 
As shown in Section~\ref{sec:QOE review}, the extracted work $\braket{W\ts{ext}}$ after completing the thermodynamic cycle reads 
\begin{align}
    \braket{W\ts{ext}}
    &=
        \braket{Q}
    =
        \braket{Q_{\mathfrak{f}_1}}
        +
        \braket{Q_{\mathfrak{f}_2}}\,,
\end{align}
where $\braket{Q}$ is the total absorbed heat from the baths during the isochoric processes, and here we additionally defined $\braket{Q_{\mathfrak{f}_j}}$ as heat exchanged during the interaction in the region $\mathfrak{f}_j$. 
To be clear, $\braket{Q_{\mathfrak{f}_1}}$ and $\braket{Q_{\mathfrak{f}_2}}$ correspond to the heat exchanged in step-\ref{step:step3} and step-\ref{step:step5}, respectively, which read $\braket{Q_{\mathfrak{f}_1}}=\Omega_1 (p_1 - p)$ and $\braket{Q_{\mathfrak{f}_2}}=\Omega_2 (p_2 - p_1)$. 
Assuming $\Omega_1 > \Omega_2$, we have seen that $p_1 - p > 0$ needs to be satisfied to extract positive work. 
This requirement implies $\braket{Q_{\mathfrak{f}_1}}>0$, meaning the qubit needs to absorb heat from the field during its first interaction in region $\mathfrak{f}_1$. 
Meanwhile, the cyclicity condition $p_2=p$ in step-\ref{step:step6} leads to $\braket{Q_{\mathfrak{f}_2}}=\Omega_2 (p - p_1) < 0$, implying that heat must be dissipated into the field during the second interaction in region $\mathfrak{f}_2$. 
In summary, the qubit must absorb heat and increase the population $p_1$ in the excited state $\ket{e}$ in step-\ref{step:step3}, followed by an adiabatic process in step-\ref{step:step4} and another isochoric process in step-\ref{step:step5} where the population decreases from $p_1$ back to $p_2(=p)$, releasing heat in the process to complete the thermodynamic cycle.

Next, we demonstrate that the delta-coupled UDW detector always absorbs heat and transitions to the excited state, as long as the population in $\ket{e}$ is less than $1/2$. 
This can be readily seen from Eqs.~\eqref{eq:p1} and \eqref{eq:p2}. 
The population $p_1$ in $\ket{e}$ after the first interaction in $\mathfrak{f}_1$ is 
\begin{align}
    p_1
    =
        \dfrac{1}{2} 
        + 
        \kako{
            p-\dfrac{1}{2}
        } 
        \nu_{\mathfrak{f}_1}\,
        \begin{cases}
            \in [ p, 1/2 ) & \text{if}~p<\dfrac{1}{2} \vspace{2mm} \\ 
            \in ( 1/2, p ] & \text{if}~p>\dfrac{1}{2}
        \end{cases} \label{eq:p after interaction without communication}
\end{align}
where we have used the fact that $\nu_{\mathfrak{f}_1} \in (0, 1]$ for a quasi-free state. 
In words, if the initial population is $p<1/2$ then an interaction governed by the delta-coupling always increases the population by absorbing heat from the field. 
Conversely, if $p>1/2$ then de-excitation and heat dissipation occur. 
For the second interaction in region $\mathfrak{f}_2$, $p_2$ is given by \eqref{eq:p2}. 
In the absence of signaling, where $E(\mathfrak{f}_1, \mathfrak{f}_2)=0$, \eqref{eq:p2} can be written as 
\begin{align}
    p_2
    &=
        \dfrac{1}{2} 
        + 
        \kako{
            p-\dfrac{1}{2}
        } 
        \nu_{\mathfrak{f}_1} \nu_{\mathfrak{f}_2} \alpha\,
        \begin{cases}
            \in [ p, 1/2 ) & \text{if}~p<\dfrac{1}{2} \vspace{2mm} \\ 
            \in ( 1/2, p ] & \text{if}~p>\dfrac{1}{2}
        \end{cases} \label{eq:p_2 after interaction without communication}
\end{align}
where 
\begin{align}
    \alpha 
    &\coloneqq
        e^{ 4 \mu (E\mathfrak{f}_1, E\mathfrak{f}_2) }
        \sin^2 
        \dfrac{\theta}{2}
        +
        e^{ -4 \mu (E\mathfrak{f}_1, E\mathfrak{f}_2) }
        \cos^2 
        \dfrac{\theta}{2} \,.
\end{align}
The ranges of $p_2$ in \eqref{eq:p_2 after interaction without communication} are due to the fact that $\nu_{\mathfrak{f}_1} \nu_{\mathfrak{f}_2} \alpha \in (0, 1]$, as given by \eqref{eq:pee pee poo poo}.

Eqs.~\eqref{eq:p after interaction without communication} and \eqref{eq:p_2 after interaction without communication} explain how the UDW detector extracts work from a field. 
First, for the positive work condition $p_1 - p > 0$ to be met, it is necessary that $p<1/2$, as the first interaction \eqref{eq:p after interaction without communication} solely increases the population. 
In this scenario, Eq.~\eqref{eq:p after interaction without communication} implies that $p_1$ also remains below $1/2$. 
Suppose the second interaction in region $\mathfrak{f}_2$ is unaffected by the first in $\mathfrak{f}_1$, i.e., no signal is exchanged. 
Then, $p_2$ is given by \eqref{eq:p_2 after interaction without communication}. 
However, since $p < 1/2$ the detector gets excited by absorbing heat from the field, conflicting with the condition $\braket{Q_{\mathfrak{f}_2}}<0$. 
Thus, in a QOE framework, a detector without signaling cannot extract work from the field because it only absorbs heat, thereby failing to close the thermodynamic cycle.

The situation changes when the detector communicates with itself from the past. 
In this case, $p_2$ is given by \eqref{eq:p2}, which is not restricted to $p_2 \in [p_1, 1/2)$. 
If this self-communication results in the detector dissipating heat, the thermodynamic cycle can be successfully closed. 
Thus, communication plays a vital role in completing the cycle.

\subsection{Example: An inertial detector extracts work from the Minkowski vacuum}\label{subsec:Example of Minkowski}

As a demonstration, consider an inertial UDW detector in $(3+1)$-dimensional Minkowski spacetime. 
We choose a massless minimally coupled scalar field [i.e., $m=0$ and $\xi=0$ in \eqref{eq:KG eq}] in the Minkowski vacuum. 
In this case, the unsmeared field operator $\hat \phi(\sx)$ at $\sx \in \mfd$ can be written as 
\begin{align}
    \hat \phi(\sx)
    &=
        \int_{\R^3} \dfrac{ \dd^3 k }{ \sqrt{ (2\pi)^3 2\kk } } 
        \kako{
            \hat a_\bk e^{ -\ii \kk t + \ii \bk \cdot \bx }
            + 
            \hat a_\bk^\dag e^{ \ii \kk t - \ii \bk \cdot \bx }
        }\,, \label{eq:Minkowski field operator}
\end{align}
with $\hat a_\bk \ket{0}=0$ for all $\bk$.

To evaluate the extracted work \eqref{eq:work extracted by delta switch}, we shall compute $\nu_{\mathfrak{f}}, E(\mathfrak{f}_1, \mathfrak{f}_2)$, and $\mu(E\mathfrak{f}_1, E\mathfrak{f}_2)$. 
As one can see from Eqs.~\eqref{eq:nu_f} and \eqref{eq:wightman and mu and causal propagator}, the spatially smeared Wightman function $\mathsf{W}(\mathfrak{f}_1, \mathfrak{f}_2)$ is a key ingredient. 
The unsmeared vacuum Wightman function from \eqref{eq:Minkowski field operator} is known to be 
\begin{align}
    \mathsf{W}(\sx, \sx')
    &=
        \braket{0|\hat \phi(\sx) \hat \phi(\sx')|0}
    =
        \int_{\R^3} \dfrac{ \dd^3 k }{ (2\pi)^3 2\kk } 
        e^{ -\ii \kk (t-t') + \ii \bk \cdot (\bx - \bx') }\,.
\end{align}
Therefore, the spatially smeared vacuum Wightman function can be written as 
\begin{align}
    \mathsf{W}(\mathfrak{f}_1, \mathfrak{f}_2)
    &=
        \int_{\R^3} \dd^3 x
        \int_{\R^3} \dd^3 x'\,
        W(\tau_1, \bx; \tau_2, \bx')
        \tilde \lambda_1 \tilde \lambda_2
        F_1(\bx) F_2(\bx') \\
    &=
        \dfrac{\tilde \lambda_1 \tilde \lambda_2}{2}
        \int_{\R^3} \dd^3 k\,
        \dfrac{ e^{ -\ii \kk \Delta \tau } }{\kk} \tilde F_1(\bk) \tilde F_2(-\bk)\,,\label{eq:smeared Wightman in Minkowski general}
\end{align}
where $\tilde \lambda_j \coloneqq \lambda_j \eta$, $\Delta \tau \coloneqq \tau_2 - \tau_1$, and 
\begin{align}
    \tilde F_j(\bk) 
    &\coloneqq
        \int_{\R^3} \dfrac{ \dd^3 x }{ \sqrt{ (2\pi)^3 } }
        F_j(\bx) e^{ \ii \bk \cdot \bx }
\end{align}
is the Fourier transform of the smearing function $F_j(\bx)$. 
Hence, we obtain $\nu_{\mathfrak{f}}$, $E(\mathfrak{f}_1, \mathfrak{f}_2)$, and $\mu(E\mathfrak{f}_1, E\mathfrak{f}_2)$ once an explicit form of the smearing function $F_j(\bx)$ is specified.

\begin{figure}[t]
\centering
\includegraphics[width=\linewidth]{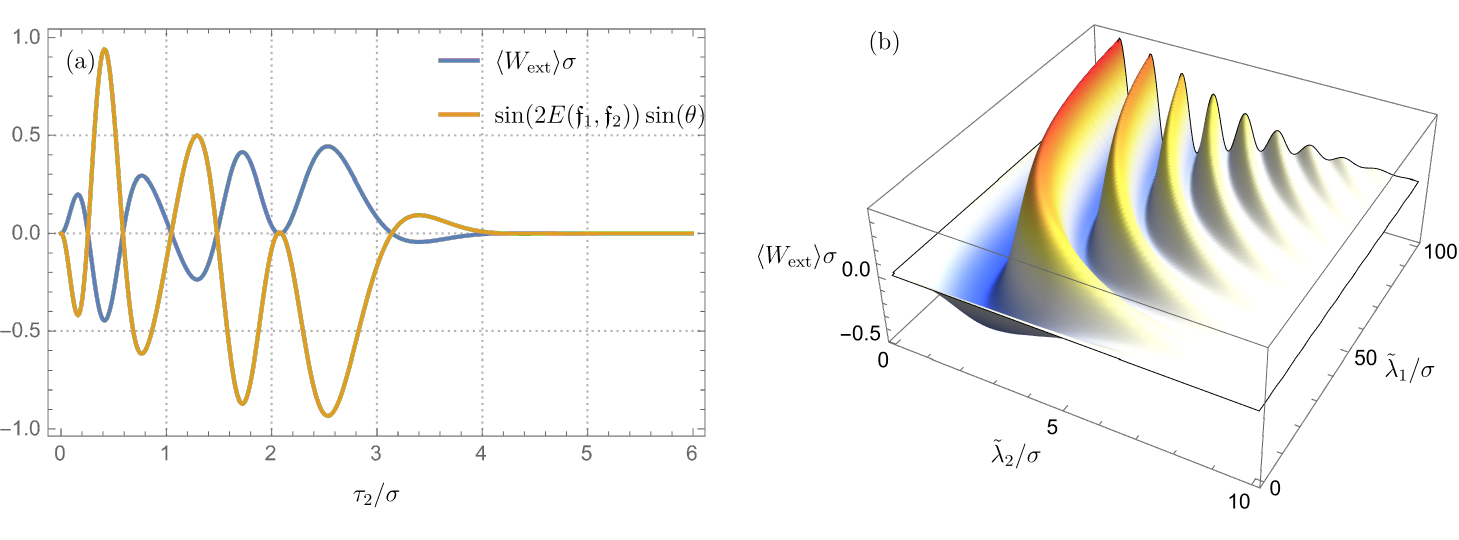}
\caption{The extracted work $\braket{W\ts{ext}}\sigma$ in units of the characteristic width $\sigma$ of the Gaussian smearing function. 
(a) Work as a function of the second interaction time $\tau_2/\sigma$ when $\tau_1/\sigma=0$, $\Omega_1 \sigma=1$, $\Omega_2 \sigma=3$, $\tilde \lambda_1 / \sigma = 100$, and $\tilde \lambda_2/\sigma = 1$. 
Negative $\braket{W\ts{ext}}\sigma$ corresponds to work done on the detetor. 
The yellow plot shows the positive work condition \eqref{eq:positive work condition for delta}. 
(b) Work as a function of the coupling constants. 
Here, we choose $\tau_1/\sigma=0$, $\Omega_1 \sigma=1$, $\Omega_2\sigma=3$, and $\tau_2/\sigma=1.5$. }
\label{fig:Minkowski example}
\end{figure}

Let us choose a Gaussian smearing function with a characteristic Gaussian width $\sigma$: 
\begin{align}
    F(\bx)
    \equiv 
    F_1(\bx)
    =
    F_2(\bx)
    =
        \dfrac{1}{ (\sqrt{\pi} \sigma)^3 }
        e^{ -|\bx|^2/\sigma^2 }\,,
\end{align}
whose Fourier transform reads 
\begin{align}
    \tilde F(\kk)
    &=
        \dfrac{ 1 }{ \sqrt{ (2\pi)^3 } }
        e^{ -\kk^2 \sigma^2/4 }\,. \label{eq:Gaussian smearing Fourier}
\end{align}
Although the Gaussian function does not have compact support, its `tails' are exponentially suppressed, allowing us to assume that the detector is effectively localized.

Inserting \eqref{eq:Gaussian smearing Fourier} into \eqref{eq:smeared Wightman in Minkowski general}, a straightforward calculation gives us \cite{GallockEntangledDetectors, Tjoa.RQC}
\begin{subequations}
    \begin{align}
        \nu_{\mathfrak{f}_j}
        &=
            e^{ -2 \mathsf{W}(\mathfrak{f}_j, \mathfrak{f}_j) }
        =
            \exp
            \kako{
                -\dfrac{ \tilde \lambda_j^2 }{ 2\pi^2 \sigma^2 }
            }\,, \label{eq:nu for Minkowski} \\
        E(\mathfrak{f}_1, \mathfrak{f}_2)
        &=
            \dfrac{1}{2}\text{Im}[ \mathsf{W}(\mathfrak{f}_1, \mathfrak{f_2}) ]
        =
            \dfrac{ \tilde \lambda_1 \tilde \lambda_2 }{ 2 \sqrt{ \pi^3 } \sigma^2 }
            \dfrac{ \Delta \tau }{ \sqrt{2} \sigma }
            \exp
            \kagikako{
                - 
                \kako{
                    \dfrac{ \Delta \tau }{ \sqrt{2} \sigma }
                }^2
            }\,, \\
        \mu(E\mathfrak{f}_1, E\mathfrak{f}_2)
        &=
            \text{Re}[ \mathsf{W}(\mathfrak{f}_1, \mathfrak{f_2}) ]
        =
            \dfrac{\tilde \lambda_1 \tilde \lambda_2}{ 4 \pi^2 \sigma^2 }
            \kagikako{
                1
                - 
                2 \dfrac{\Delta \tau}{ \sqrt{2} \sigma }
                D^+
                \kako{
                    \dfrac{\Delta \tau}{ \sqrt{2} \sigma }
                }
            }\,,
    \end{align}
\end{subequations}
where $D^+(x)$ is the Dawson function given by 
\begin{align}
    D^+(x)
    \coloneqq
        \dfrac{\sqrt{\pi}}{2} e^{-x^2} \erfi(x)
\end{align}
with the imaginary error function $\erfi(x)$.

After substituting these expressions into the extracted work \eqref{eq:work extracted by delta switch}, the results are shown in Figure~\ref{fig:Minkowski example}. 
In Figure~\ref{fig:Minkowski example}(a), we depict the dependence of the second interaction's timing, $\tau_2/\sigma$, in units of the characteristic Gaussian width $\sigma$, given $\tau_1/\sigma=0$, $\Omega_1 \sigma=1$, $\Omega_2 \sigma=3$, $\tilde \lambda_1 /\sigma=100$, and $\tilde \lambda_2 /\sigma=1$. 
As discussed in Section~\ref{subsec: observation}, the detector can extract positive work from the vacuum state of the field instantaneously as long as the first interaction affects the second one. 
In fact, as one can see from the positive work condition depicted as the yellow curve in \ref{fig:Minkowski example}(a), we have $\braket{W\ts{ext}}\sigma=0$ once the detector cannot receive the signal from the first interaction (i.e., when $\tau_2/\sigma > 4$).

Figure~\ref{fig:Minkowski example}(b) shows the coupling-dependence of the work $\braket{W\ts{ext}}\sigma$ when $\tau_1/\sigma=0$, $\Omega_1 \sigma=1$, $\Omega_2\sigma=3$, and $\tau_2/\sigma=1.5$. 
We see that a pair $(\tilde \lambda_1, \tilde \lambda_2)$ must be chosen carefully to satisfy the positive work condition. 
In particular, the first coupling $\tilde \lambda_1$ is preferred to be strong while the second one $\tilde \lambda_2$ should be much weaker. 
This is due to the fact that $\braket{W\ts{ext}} \to 0$ as $\nu_{\mathfrak{f}_2} \to 0$ in \eqref{eq:work extracted by delta switch}, which is the case for $\tilde \lambda_2/\sigma \gg 1$ in \eqref{eq:nu for Minkowski}.

\section{Conclusion}\label{sec:conclusion}

We have examined the relativistic QOE using the UDW particle detector model as a working substance interacting with a quantum scalar field. 
Here, the background spacetime is assumed to be an arbitrary globally hyperbolic spacetime and the state of the field is considered to be a quasi-free state such as vacuum and thermal states. 
By adopting algebraic quantum field theoretical formulation, we were able to consider such a very general setting.

We were particularly interested in the so-called delta-coupling model of the UDW detector for two reasons. 
First, the delta-coupling---which is implemented by writing the time dependence of the interaction as Dirac's delta distribution---enables us to employ a non-perturbative analysis, thereby a strong coupling can be considered. 
Secondly, it allows us to examine if a detector can extract work from a quantum field by means of instantaneous interactions.

Under these assumptions, we have derived the extracted work \eqref{eq:work extracted by delta switch} and its positive work condition \eqref{eq:positive work condition for delta} non-perturbatively and found that it is indeed possible to extract work instantaneously from a relativistic QOE as long as the detector receives the signal from its past self. 
That is, the detector must receive a signal during its second interaction that was transmitted by the quantum field from the first interaction. 
This is because each interaction without signaling only absorbs heat from the field, thus failing to complete the thermodynamic cycle. 
On the other hand, self-communication may assist the detector in releasing heat back into the field, thereby the completion of the cycle. 
As an example, we considered a scenario where the detector is at rest in flat spacetime. 
We showed that the detector is able to extract work from the Minkowski vacuum due to the signaling effect. 

Since our interaction model is instantaneous, one might question its practicality in the context of a quantum heat engine. 
Indeed, the necessity for the detector to receive a signal during the second isochoric process suggests that the time interval between the two isochoric processes must be exceptionally short, thereby necessitating an immediate adiabatic process in between. 
Nevertheless, our results highlight the crucial role of the back-reaction effect from the reservoir in quantum heat engines. 
That is, the influence of the first isochoric process on the second through a quantum field (serving as the bath in this context) makes it possible for the detector to extract work.


As an extension to this paper, it would be interesting to include a quantum field in the Gaussian states with a non-vanishing one-point Wightman function such as a coherent state.

\textbf{Note:} A related study was recently conducted in \cite{kollas2023exactly}. 
In this study, the authors consider a relativistic QOE utilizing a delta-coupled UDW detector in Minkowski spacetime to extract work from a quantum field in a thermal state.

\appendix

\section{Second interaction}\label{app:second interaction calculation}

We provide a detailed calculation for the second interaction. 
We first assume the initial field state $\hat \rho_{\phi,0}$ to be arbitrary and consider a quasi-free state at the end.

The total density operator $\hat \rho\ts{tot,2}$ right after the second interaction is given by 
\begin{align}
    \hat \rho\ts{tot,2}
    &=
        \hat U\ts{I}^{(2)} \hat U\ts{I}^{(1)}
        \hat \rho\ts{tot,0}
        \hat U\ts{I}^{(1)\dagger} \hat U\ts{I}^{(2)\dagger} \notag \\
    &=
        \hat \rho\ts{D,0} \otimes \hat C_2 \hat C_1 \hat \rho_{\phi,0} \hat C_1 \hat C_2
        - \ii \hat \mu_2 \hat \rho\ts{D,0} \otimes \hat S_2 \hat C_1 \hat \rho_{\phi,0} \hat C_1 \hat C_2
        + \ii \hat \rho\ts{D,0} \hat \mu_2 \otimes \hat C_2 \hat C_1 \hat \rho_{\phi,0} \hat C_1 \hat S_2 \notag \\
        &\quad
        + 
        \hat \mu_2 \hat \rho\ts{D,0} \hat \mu_2 \otimes \hat S_2 \hat C_1 \hat \rho_{\phi,0} \hat C_1 \hat S_2
        +
        \hat \mu_1 \hat \rho\ts{D,0} \hat \mu_1 \otimes \hat C_2 \hat S_1 \hat \rho_{\phi,0} \hat S_1 \hat C_2 \notag \\
        &\quad
        -\ii \hat \mu_2 \hat \mu_1 \hat \rho\ts{D,0} \hat \mu_1 \otimes \hat S_2 \hat S_1 \hat \rho_{\phi,0} \hat S_1 \hat C_2 
        +
        \ii \hat \mu_1 \hat \rho\ts{D,0} \hat \mu_1 \hat \mu_2 \otimes \hat C_2 \hat S_1 \hat \rho_{\phi,0} \hat S_1 \hat S_2 \notag \\
        &\quad
        +
        \hat \mu_2 \hat \mu_1 \hat \rho\ts{D,0} \hat \mu_1 \hat \mu_2 \otimes \hat S_2 \hat S_1 \hat \rho_{\phi,0} \hat S_1 \hat S_2
        -
        \ii \hat \mu_1 \hat \rho\ts{D,0} \otimes \hat C_2 \hat S_1 \hat \rho_{\phi,0} \hat C_1 \hat C_2 \notag \\
        &\quad
        -
        \hat \mu_2 \hat \mu_1 \hat \rho\ts{D,0} \otimes \hat S_2 \hat S_1 \hat \rho_{\phi,0} \hat C_1 \hat C_2
        +
        \hat \mu_1 \hat \rho\ts{D,0} \hat \mu_2 \otimes \hat C_2 \hat S_1 \hat \rho_{\phi,0} \hat C_1 \hat S_2 \notag \\
        &\quad
        -
        \ii \hat \mu_2 \hat \mu_1 \hat \rho\ts{D,0} \hat \mu_2 \otimes \hat S_2 \hat S_1 \hat \rho_{\phi,0} \hat C_1 \hat S_2 
        +
        \ii \hat \rho\ts{D,0} \hat \mu_1 \otimes \hat C_2 \hat C_1 \hat \rho_{\phi,0} \hat S_1 \hat C_2 \notag \\
        &\quad
        +
        \hat \mu_2 \hat \rho\ts{D,0} \hat \mu_1 \otimes \hat S_2 \hat C_1 \hat \rho_{\phi,0} \hat S_1 \hat C_2
        - 
        \hat \rho\ts{D,0} \hat \mu_1 \hat \mu_2 \otimes \hat C_2 \hat C_1 \hat \rho_{\phi,0} \hat S_1 \hat S_2 \notag \\
        &\quad
        + 
        \ii \hat \mu_2 \hat \rho\ts{D,0} \hat \mu_1 \hat \mu_2 \otimes \hat S_2 \hat C_1 \hat \rho_{\phi,0} \hat S_1 \hat S_2\,,
\end{align}
and the final density matrix for the detector, $\hat \rho\ts{D,2}$, can be obtained by tracing out the field degree of freedom: 
\begin{align}
    \hat \rho\ts{D,2}
    &=
        \Tr_\phi [ \hat \rho\ts{tot,2} ] \notag \\
    &=
        \hat \rho\ts{D,0}  \omega(C_1 C_2^2 C_1) 
        + 
        \ii [\hat \rho\ts{D,0}, \hat \mu_2] \omega(C_1 S_2 C_2 C_1) 
        + 
        \hat \mu_2 \hat \rho\ts{D,0} \hat \mu_2 \omega(C_1 S_2^2 C_1) \notag \\
        &\quad
        +
        \hat \mu_1 \hat \rho\ts{D,0} \hat \mu_1 \omega(S_1 C_2^2 S_1) 
        +
        \ii [\hat \mu_1 \hat \rho\ts{D,0} \hat \mu_1, \hat \mu_2] \omega(S_1 S_2 C_2 S_1) 
        +
        \hat \mu_2 \hat \mu_1 \hat \rho\ts{D,0} \hat \mu_1 \hat \mu_2 \omega(S_1 S_2^2 S_1) \notag \\
        &\quad
        -
        \ii \hat \mu_1 \hat \rho\ts{D,0} \omega(C_1 C_2^2 S_1)
        +
        [\hat \mu_1 \hat \rho\ts{D,0}, \hat \mu_2] \omega(C_1 S_2 C_2 S_1) 
        -
        \ii \hat \mu_2 \hat \mu_1 \hat \rho\ts{D,0} \hat \mu_2 \omega(C_1 S_2^2 S_1) \notag \\
        &\quad
        +
        \ii \hat \rho\ts{D,0} \hat \mu_1 \omega(S_1 C_2^2 C_1) 
        - 
        [\hat \rho\ts{D,0}\hat \mu_1, \hat \mu_2] \omega(S_1 S_2 C_2 C_1) 
        + 
        \ii \hat \mu_2 \hat \rho\ts{D,0} \hat \mu_1 \hat \mu_2 \omega(S_1 S_2^2 C_1)\,,
        \label{eq:rhoD2 1}
\end{align}
where we used $[\hat C_j, \hat S_j]=0$ at the end.

We now calculate each algebraic state. 
As we did in Section~\ref{subsec:first interaction}, we write $S_j$ and $C_j$ in terms of the elements of the Weyl algebra, Eq.~\eqref{eq:S and C in terms of Weyl}. 
Each algebraic state in \eqref{eq:rhoD2 1} takes the form 
\begin{align}
    &\omega 
    \kako{
        \dfrac{ W_{\mathfrak{f}_1} +(-1)^r W_{\mathfrak{f}_1}^* }{2\ii^r}
        \dfrac{ W_{\mathfrak{f}_2} +(-1)^s W_{\mathfrak{f}_2}^* }{2\ii^s}
        \dfrac{ W_{\mathfrak{f}_2} +(-1)^u W_{\mathfrak{f}_2}^* }{2\ii^u}
        \dfrac{ W_{\mathfrak{f}_1} +(-1)^v W_{\mathfrak{f}_1}^* }{2\ii^v}
    } \,,
\end{align}
where $r, s, u, v \in \{ 1,\, -1 \}$. 
A straightforward calculation ends up with 
\begin{align}
    &\omega 
    \kako{
        \dfrac{ W_{\mathfrak{f}_1} +(-1)^r W_{\mathfrak{f}_1}^* }{2\ii^r}
        \dfrac{ W_{\mathfrak{f}_2} +(-1)^s W_{\mathfrak{f}_2}^* }{2\ii^s}
        \dfrac{ W_{\mathfrak{f}_2} +(-1)^u W_{\mathfrak{f}_2}^* }{2\ii^u}
        \dfrac{ W_{\mathfrak{f}_1} +(-1)^v W_{\mathfrak{f}_1}^* }{2\ii^v}
    } \notag \\
    &=
        \dfrac{1}{16 \ii^{r+s+u+v}} \notag \\
        &\times 
        \Bigkagikako{
            \omega(W_{2\mathfrak{f}_1 + 2 \mathfrak{f}_2}) 
            +
            (-1)^r 
            e^{2\ii E(\mathfrak{f}_1, \mathfrak{f}_2)} 
            \omega(W_{2\mathfrak{f}_2}) 
            +
            (-1)^v 
            e^{-2\ii E(\mathfrak{f}_1, \mathfrak{f}_2)} 
            \omega(W_{2\mathfrak{f}_2}) 
            +
            (-1)^{r+v} 
            \omega(W_{2\mathfrak{f}_1 - 2 \mathfrak{f}_2}^*) \notag \\
            &\quad
            +
            (-1)^s 
            \omega(W_{2\mathfrak{f}_1}) 
            +
            (-1)^{u} 
            \omega(W_{2\mathfrak{f}_1}) 
            +
            (-1)^{r+s}
            +
            (-1)^{r+u}
            +
            (-1)^{s+v} 
            +
            (-1)^{u+v} \notag \\
            &\quad
            +
            (-1)^{r+s+v} 
            \omega(W_{2\mathfrak{f}_1}^*) 
            +
            (-1)^{r+u+v} 
            \omega(W_{2\mathfrak{f}_1}^*)  
            +
            (-1)^{s+u} 
            \omega(W_{2\mathfrak{f}_1 - 2 \mathfrak{f}_2}) \notag \\
            &\quad
            +
            (-1)^{r+s+u} 
            e^{-2\ii E(\mathfrak{f}_1, \mathfrak{f}_2)}
            \omega(W_{2\mathfrak{f}_2}^*) 
            +
            (-1)^{s+u+v} 
            e^{2\ii E(\mathfrak{f}_1, \mathfrak{f}_2)}
            \omega(W_{2\mathfrak{f}_2}^*) \notag \\
            &\quad
            +
            (-1)^{r+s+u+v} 
            \omega(W_{2\mathfrak{f}_1 + 2 \mathfrak{f}_2}^*)
        }\,.
\end{align}
Here, we used the fact that $\omega: \mathcal{W}(\mfd) \to \C$ is a linear map, and utilized an identity 
\begin{align}
    &W_{a \mathfrak{f}_1} W_{b \mathfrak{f}_2} W_{c \mathfrak{f}_2} W_{d \mathfrak{f}_1}
    =
        e^{-\ii (a-d)(b+c) E(\mathfrak{f}_1,\mathfrak{f}_2)/2}
        W_{(a+d) \mathfrak{f}_1 + (b+c)\mathfrak{f}_2}\,, 
\end{align}
for $a,b,c,d \in \R$, which can be derived from the Weyl relations \eqref{eq:Weyl relations}.

So far, we have assumed that the initial state of the field is arbitrary. 
We now specify $\hat \rho_{\phi,0}$ to be a quasi-free state \eqref{eq:quasifree state}, and define
\begin{align}
    \nu_{f_j} 
    \equiv
        \omega(W_{2\mathfrak{f}_j})
    =
        e^{-2\mathsf{W}(\mathfrak{f}_j, \mathfrak{f}_j)} \in (0,1]\,,
\end{align}
which leads to
\begin{align}
    &\omega 
    \kako{
        \dfrac{ W_{\mathfrak{f}_1} +(-1)^r W_{\mathfrak{f}_1}^* }{2\ii^r}
        \dfrac{ W_{\mathfrak{f}_2} +(-1)^s W_{\mathfrak{f}_2}^* }{2\ii^s}
        \dfrac{ W_{\mathfrak{f}_2} +(-1)^u W_{\mathfrak{f}_2}^* }{2\ii^u}
        \dfrac{ W_{\mathfrak{f}_1} +(-1)^v W_{\mathfrak{f}_1}^* }{2\ii^v}
    } \notag \\
    &=
        \dfrac{1}{16 \ii^{r+s+u+v}} \notag \\
        &\times 
        \Bigkako{
            [1 + (-1)^{r+s+u+v} ]
            \nu_{\mathfrak{f}_1+\mathfrak{f}_2}
            +
            [
                (-1)^r + (-1)^{s+u+v} 
            ]
            e^{2\ii E(\mathfrak{f}_1, \mathfrak{f}_2)} 
            \nu_{\mathfrak{f}_2} \notag \\
            &\quad
            +
            [
                (-1)^v + (-1)^{r+s+u} 
            ]
            e^{-2\ii E(\mathfrak{f}_1, \mathfrak{f}_2)} 
            \nu_{\mathfrak{f}_2}
            +
            [
                (-1)^{r+v} + (-1)^{s+u} 
            ]
            \nu_{\mathfrak{f}_1-\mathfrak{f}_2} \notag \\
            &\quad
            +
            [
                (-1)^s + (-1)^{u} + (-1)^{r+s+v} + (-1)^{r+u+v} 
            ]
            \nu_{\mathfrak{f}_1} \notag \\
            &\quad
            +
            (-1)^{r+s}
            +
            (-1)^{r+u}
            +
            (-1)^{s+v}
            +
            (-1)^{u+v}
        }\,.
\end{align}
One can directly check that if one of $r,s,u,v$ has a different value than the rest of the three then the whole term vanishes. 
This corresponds to, for instance, $\omega(C_1 S_2 C_2 C_1)=0$ and so on. 
Therefore, if the field is initially in a quasi-free state, the final density operator of the detector \eqref{eq:rhoD2 1} reads
\begin{align}
    \hat \rho\ts{D,2}
    &=
        \hat \rho\ts{D,0}  \omega(C_1 C_2^2 C_1) 
        + 
        \hat \mu_2 \hat \rho\ts{D,0} \hat \mu_2 \omega(C_1 S_2^2 C_1) 
        +
        \hat \mu_1 \hat \rho\ts{D,0} \hat \mu_1 \omega(S_1 C_2^2 S_1) \notag \\
        &\quad
        +
        \hat \mu_2 \hat \mu_1 \hat \rho\ts{D,0} \hat \mu_1 \hat \mu_2 \omega(S_1 S_2^2 S_1) 
        +
        [\hat \mu_1 \hat \rho\ts{D,0}, \hat \mu_2] \omega(C_1 S_2 C_2 S_1) 
        - 
        [\hat \rho\ts{D,0}\hat \mu_1, \hat \mu_2] \omega(S_1 S_2 C_2 C_1)\,,
\end{align}
where each algebraic state is explicitly written as 
\begin{subequations}
    \begin{align}
        &\omega(C_1 C_2^2 C_1)
        =
            \dfrac{1}{16}
            \Bigkagikako{
                4 
                + 4 \nu_{\mathfrak{f}_1}
                + 2 \nu_{\mathfrak{f}_1-\mathfrak{f}_2}
                + 2 \nu_{\mathfrak{f}_1+\mathfrak{f}_2}
                + 4 \cos (2E(\mathfrak{f}_1, \mathfrak{f}_2)) \nu_{\mathfrak{f}_2}
            }\,,\\
        &\omega(C_1 S_2^2 C_1)
        =
            \dfrac{1}{16}
            \Bigkagikako{
                4 
                + 4 \nu_{\mathfrak{f}_1}
                - 2 \nu_{\mathfrak{f}_1-\mathfrak{f}_2}
                - 2 \nu_{\mathfrak{f}_1+\mathfrak{f}_2}
                - 4 \cos (2E(\mathfrak{f}_1, \mathfrak{f}_2)) \nu_{\mathfrak{f}_2}
            }\,,\\
        &\omega(S_1 C_2^2 S_1)
        =
            \dfrac{1}{16}
            \Bigkagikako{
                4 
                - 4 \nu_{\mathfrak{f}_1}
                - 2 \nu_{\mathfrak{f}_1-\mathfrak{f}_2}
                - 2 \nu_{\mathfrak{f}_1+\mathfrak{f}_2}
                + 4 \cos (2E(\mathfrak{f}_1, \mathfrak{f}_2)) \nu_{\mathfrak{f}_2}
            }\,,\\
        &\omega(S_1 S_2^2 S_1)
        =
            \dfrac{1}{16}
            \Bigkagikako{
                4 
                - 4 \nu_{\mathfrak{f}_1}
                + 2 \nu_{\mathfrak{f}_1-\mathfrak{f}_2}
                + 2 \nu_{\mathfrak{f}_1+\mathfrak{f}_2}
                - 4 \cos (2E(\mathfrak{f}_1, \mathfrak{f}_2)) \nu_{\mathfrak{f}_2}
            }\,,\\
        &\omega(C_1 C_2 S_2 S_1)
        =
            \omega(C_1 S_2 C_2 S_1)
        =
            \dfrac{1}{16}
            \Bigkagikako{
                2 \nu_{\mathfrak{f}_1-\mathfrak{f}_2}
                - 2 \nu_{\mathfrak{f}_1+\mathfrak{f}_2}
                - 4\ii \sin (2E(\mathfrak{f}_1, \mathfrak{f}_2)) \nu_{\mathfrak{f}_2}
            }\,,\\
        &\omega(S_1 C_2 S_2 C_1)
        =
            \omega(S_1 S_2 C_2 C_1)
        =
            \dfrac{1}{16}
            \Bigkagikako{
                2 \nu_{\mathfrak{f}_1-\mathfrak{f}_2}
                - 2 \nu_{\mathfrak{f}_1+\mathfrak{f}_2}
                + 4\ii \sin (2E(\mathfrak{f}_1, \mathfrak{f}_2)) \nu_{\mathfrak{f}_2}
            }\,.
    \end{align}
\end{subequations}
One can further simplify these by using the following: 
\begin{subequations}
    \begin{align}
        \nu_{\mathfrak{f}_1-\mathfrak{f}_2} + \nu_{\mathfrak{f}_1+\mathfrak{f}_2}
        &=
            2 \nu_{\mathfrak{f}_1} \nu_{\mathfrak{f}_2} \cosh[ 4\mu(E\mathfrak{f}_1, E\mathfrak{f}_2) ]\,,\\
        \nu_{\mathfrak{f}_1-\mathfrak{f}_2} - \nu_{\mathfrak{f}_1+\mathfrak{f}_2}
        &=
            2 \nu_{\mathfrak{f}_1} \nu_{\mathfrak{f}_2} \sinh[ 4\mu(E\mathfrak{f}_1, E\mathfrak{f}_2) ]\,.
    \end{align}
\end{subequations}
Here, $\mu(E\mathfrak{f}_1, E\mathfrak{f}_2) \in \R$ is given in \eqref{eq:wightman and mu and causal propagator}. 
This can be proven as follows. 
First, we recall that $\nu_\mathfrak{f} = \omega (W_{2\mathfrak{f}})$. 
For quasi-free states, we can write this as 
\begin{align}
    \nu_\mathfrak{f}
    &=
        \omega(W_{2\mathfrak{f}})
    =
        e^{ -2\mu(E\mathfrak{f}, E\mathfrak{f}) }\,.
\end{align}
Also, from the linearity of the causal propagator operator, $E(f+f')= Ef + Ef'$, we have 
\begin{align}
    \nu_{\mathfrak{f}_1 - \mathfrak{f}_2} + \nu_{\mathfrak{f}_1 + \mathfrak{f}_2}
    &=
        e^{ -2\mu(E(\mathfrak{f}_1-\mathfrak{f}_2), E(\mathfrak{f}_1-\mathfrak{f}_2)) }
        +
        e^{ -2\mu(E(\mathfrak{f}_1+\mathfrak{f}_2), E(\mathfrak{f}_1+\mathfrak{f}_2)) } \notag \\
    &=
        e^{-2\mu(E\mathfrak{f}_1, E\mathfrak{f}_1)}
        e^{2\mu(E\mathfrak{f}_1, E\mathfrak{f}_2)}
        e^{2\mu(E\mathfrak{f}_1, E\mathfrak{f}_2)}
        e^{-2\mu(E\mathfrak{f}_2, E\mathfrak{f}_2)} \notag \\
        &\quad 
        +
        e^{-2\mu(E\mathfrak{f}_1, E\mathfrak{f}_1)}
        e^{-2\mu(E\mathfrak{f}_1, E\mathfrak{f}_2)}
        e^{-2\mu(E\mathfrak{f}_1, E\mathfrak{f}_2)}
        e^{-2\mu(E\mathfrak{f}_2, E\mathfrak{f}_2)} \notag \\
    &=
        \nu_{\mathfrak{f}_1} \nu_{\mathfrak{f}_2} e^{ 4\mu(E\mathfrak{f}_1, E\mathfrak{f}_2) }
        +
        \nu_{\mathfrak{f}_1} \nu_{\mathfrak{f}_2} e^{ -4\mu(E\mathfrak{f}_1, E\mathfrak{f}_2) } \notag \\
    &=
        2 \nu_{\mathfrak{f}_1} \nu_{\mathfrak{f}_2} \cosh[ 4\mu(E\mathfrak{f}_1, E\mathfrak{f}_2) ]\,.
\end{align}
Then the aforementioned algebraic states are reduced to 
\begin{subequations}
    \begin{align}
        &\omega(C_1 C_2^2 C_1)
        =
            \dfrac{1}{4}
            \Bigkagikako{
                1 
                + \nu_{\mathfrak{f}_1}
                + \nu_{\mathfrak{f}_1} \nu_{\mathfrak{f}_2} \cosh[ 4\mu(E\mathfrak{f}_1, E\mathfrak{f}_2) ]
                + \nu_{\mathfrak{f}_2} \cos (2E(\mathfrak{f}_1, \mathfrak{f}_2)) 
            }\,,\\
        &\omega(C_1 S_2^2 C_1)
        =
            \dfrac{1}{4}
            \Bigkagikako{
                1 
                + \nu_{\mathfrak{f}_1}
                - \nu_{\mathfrak{f}_1} \nu_{\mathfrak{f}_2} \cosh[ 4\mu(E\mathfrak{f}_1, E\mathfrak{f}_2) ]
                - \nu_{\mathfrak{f}_2} \cos (2E(\mathfrak{f}_1, \mathfrak{f}_2)) 
            }\,,\\
        &\omega(S_1 C_2^2 S_1)
        =
            \dfrac{1}{4}
            \Bigkagikako{
                1 
                - \nu_{\mathfrak{f}_1}
                - \nu_{\mathfrak{f}_1} \nu_{\mathfrak{f}_2} \cosh[ 4\mu(E\mathfrak{f}_1, E\mathfrak{f}_2) ]
                + \nu_{\mathfrak{f}_2} \cos (2E(\mathfrak{f}_1, \mathfrak{f}_2)) 
            }\,,\\
        &\omega(S_1 S_2^2 S_1)
        =
            \dfrac{1}{4}
            \Bigkagikako{
                1 
                - \nu_{\mathfrak{f}_1}
                + \nu_{\mathfrak{f}_1} \nu_{\mathfrak{f}_2} \cosh[ 4\mu(E\mathfrak{f}_1, E\mathfrak{f}_2) ]
                - \nu_{\mathfrak{f}_2} \cos (2E(\mathfrak{f}_1, \mathfrak{f}_2)) 
            }\,,\\
        &\omega(C_1 C_2 S_2 S_1)
        =
            \omega(C_1 S_2 C_2 S_1)
        =
            \dfrac{\nu_{\mathfrak{f}_2}}{4}
            \Bigkagikako{
                \nu_{\mathfrak{f}_1} \sinh[ 4\mu(E\mathfrak{f}_1, E\mathfrak{f}_2) ]
                - \ii \sin (2E(\mathfrak{f}_1, \mathfrak{f}_2)) 
            }\,,\\
        &\omega(S_1 C_2 S_2 C_1)
        =
            \omega(S_1 S_2 C_2 C_1)
        =
            \dfrac{\nu_{\mathfrak{f}_2}}{4}
            \Bigkagikako{
                \nu_{\mathfrak{f}_1} \sinh[ 4\mu(E\mathfrak{f}_1, E\mathfrak{f}_2) ]
                + \ii \sin (2E(\mathfrak{f}_1, \mathfrak{f}_2)) 
            }\,.
    \end{align}
\end{subequations}

\acknowledgments{KGY is thankful to Erickson Tjoa for reviewing this manuscript and Adam Teixid\'{o}-Bonfill for a helpful comment.}

\bibliography{ref}
\bibliographystyle{JHEP}

\end{document}